\documentclass[sigconf]{acmart}
\AtBeginDocument{%
  \providecommand\BibTeX{{%
    \normalfont B\kern-0.5em{\scshape i\kern-0.25em b}\kern-0.8em\TeX}}}

\copyrightyear{2026}
\acmYear{2026}
\setcopyright{cc}
\setcctype{by-nc-nd}
\acmConference[CHI '26]{Proceedings of the 2026 CHI Conference on Human Factors in Computing Systems}{April 13--17, 2026}{Barcelona, Spain}
\acmBooktitle{Proceedings of the 2026 CHI Conference on Human Factors in Computing Systems (CHI '26), April 13--17, 2026, Barcelona, Spain}
\acmPrice{}
\acmDOI{10.1145/3772318.3790707}
\acmISBN{979-8-4007-2278-3/2026/04}

\usepackage{mydefs,acmart-taps}
\usepackage{amsmath}
\usepackage{graphicx}
\usepackage{cleveref}
\usepackage{subcaption}
\usepackage{multirow}   
\usepackage{xspace}
\usepackage{enumitem}

\definecolor{boxheader}{gray}{0.35}
\definecolor{boxbg}{gray}{0.97}
\definecolor{boxborder}{gray}{0.35}

\newcommand{\ie}{i.e.{\xspace}}
\newcommand{\eg}{e.g.{\xspace}}
\newcommand{\vs}{vs.{\xspace}}
\newcommand{\qt}[1]{\textit{``#1''}}
\newcommand{\pqt}[2]{\textit{``#1''}{\,}{\small-#2}}

\newcommand{\iqr}[1]{$IQR=#1$}
\newcommand{\md}[1]{$MD=#1$}
\newcommand{\etal}{{\xspace}et~al.{\xspace}}

\begin{document}

\newcommand{\xinyu}[1]{\authorcomment{RED}{Xinyu}{#1}}
\newcommand{\rubaiat}[1]{\authorcomment{BLUE}{Rubaiat}{#1}}
\newcommand{\liyi}[1]{\authorcomment{PURPLE}{Li-Yi}{#1}}
\newcommand{\cherry}[1]{\authorcomment{GREEN}{Cherry}{#1}}
\newcommand{\jz}[1]{\authorcomment{CYAN}{Jian}{#1}}


\raggedbottom

\newcommand{\sysname}{Notational Animating}
\newcommand{\conceptname}{\textit{notational animating}}

\title[\sysname{}]{\sysname{}: An Interactive Approach to Creating and Editing Animation Keyframes}

\author{Xinyu Shi}
\affiliation{%
  \institution{University of Waterloo}
  \city{Waterloo}
  \state{ON}
  \country{Canada}
}
\email{xinyu.shi@uwaterloo.ca}

\author{Li-Yi Wei}
\affiliation{%
  \institution{Adobe Research}
  \city{San Jose}
  \state{CA}
  \country{United States}
}
\email{lwei@adobe.com}

\author{Nanxuan Zhao}
\affiliation{%
  \institution{Adobe Research}
  \city{San Jose}
  \state{CA}
  \country{United States}
}
\email{nanxuanz@adobe.com}

\author{Jian Zhao}
\affiliation{%
  \institution{University of Waterloo}
  \city{Waterloo}
  \state{ON}
  \country{Canada}
}
\email{jianzhao@uwaterloo.ca}

\author{Rubaiat Habib Kazi}
\affiliation{%
  \institution{Adobe Research}
  \city{Seattle}
  \state{WA}
  \country{United States}
}
\email{rhabib@adobe.com}

\begin{teaserfigure}
\centering
  \includegraphics[width=\linewidth]{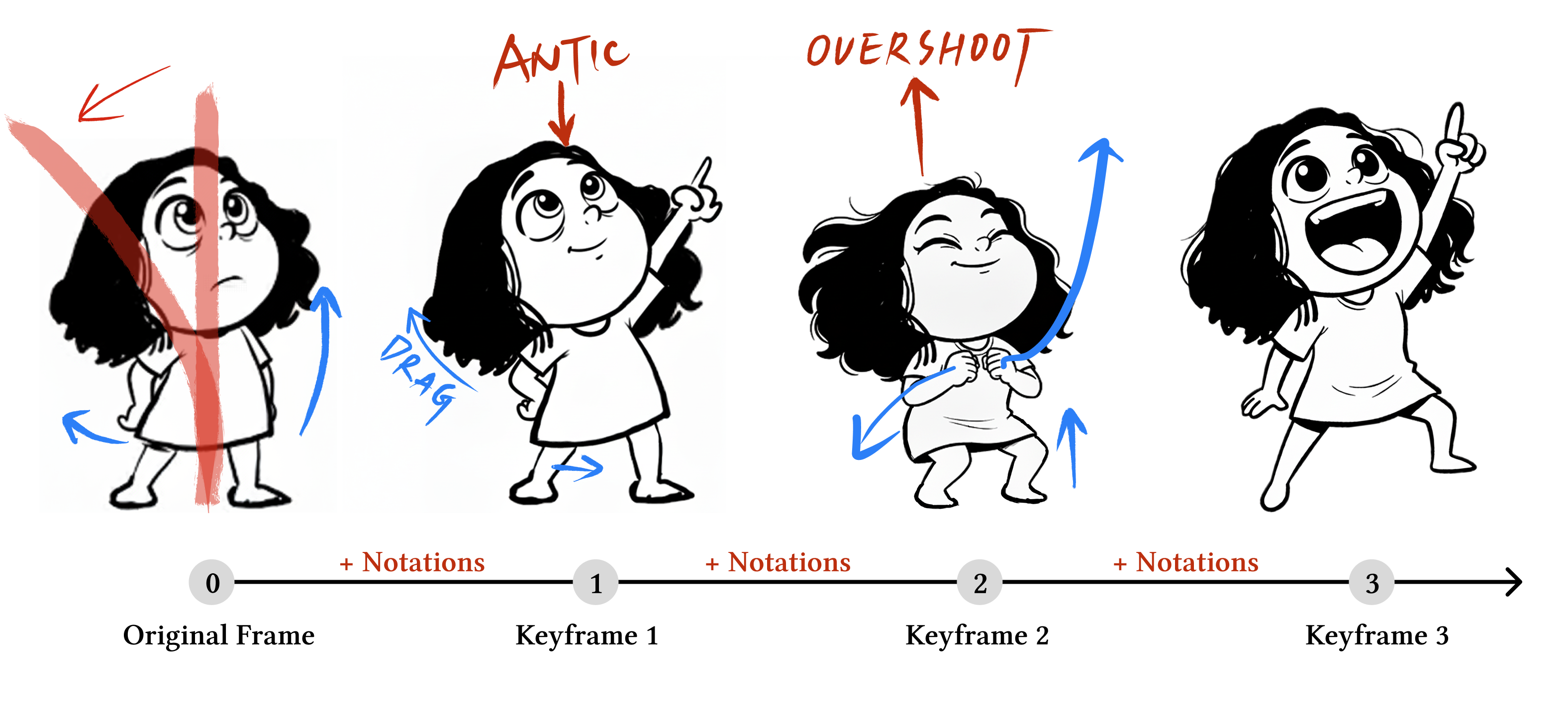}
  \vspace{-10mm}
  \caption{A usage example with \conceptname{}: starting from a static drawing (left), the animator sketches high-level notations such as arrows, strokes, and text labels. The system interprets these notations and then generates the next keyframe. By iterating this loop (add notations and generate), the artist progressively authors and edits a sequence of keyframes (1–3). }
  \Description{A four-panel storyboard shows how ``notational animating'' works. Starting from an original static drawing of a cartoon character (panel 0), the animator draws red strokes and blue curved arrows on top of the frame to indicate intended motion. In Keyframe 1, the character begins raising an arm and looking upward; handwritten labels such as ``ANTIC'' (anticipation) and ``DRAG'' appear alongside arrows that indicate direction and timing. In Keyframe 2, the character is in a compressed pose with more blue arrows showing overlapping movement, and a label ``OVERSHOOT'' indicates exaggerated motion past the target. Keyframe 3 shows the resulting energetic pose with the arm fully extended and the character smiling. A timeline along the bottom (0 -> 3) marks each keyframe, with ``+ Notations'' between frames to show the repeated loop of adding annotations and generating the next keyframe.}
  \label{fig:teaser}
\end{teaserfigure}

\begin{abstract}

We introduce the concept of \conceptname{}, an interaction paradigm for animation authoring where users sketch high-level notations over static drawings to indicate intended motions, which are then interpreted by automatic methods (\eg, GenAI models) to generate animation keyframes.
Sketched notations have long served as cognitive instruments for animators, capturing forces, poses,  dynamics, paths, and other animation features.
However, such notations are often \emph{contextual, ambiguous, and combinational} based on our analysis of 135 real-world sketches.
To facilitate interpretation, we first formalize these notations into a structured animation representation (\ie, source, path, and target).
We then built an animation authoring system that translates high-level notations into the formalized intended animation, provides dynamic UI widgets for fine-grained parameter control, and establishes a closed feedback loop to resolve ambiguity.
Finally, through a preliminary study with animators, we assess the usability of \conceptname{}, reflect its affordance, and identify its contexts of use.

\end{abstract}



\maketitle
\section{Introduction}

Expressive animation, distinct from physics-based simulation and live action films, aims to depict imagined worlds that are not necessarily realistic yet believable and compelling~\cite{thomas1981illusion}. 
The craft of expressive animation has evolved from classical flipbook-style hand drawings to computer-aided direct manipulations, and now towards generative AI-assisted authoring based on natural language inputs.
However, the tacit knowledge of how to give a performance with movement, weight, timing, and emotion is long internalized and conveyed through visual language by artists~\cite{Williams:2009:AnimatorSurvivalKit};
natural language alone is too ambiguous to capture such nuanced intent.

Drawing on animators' traditional practice, we envision a future of animation authoring where sketched notations serve as the \emph{reified user intention} \cite{beaudouin2000instrumental, Riche:2025:AI-Instruments} to create and edit keyframes that can be subsequently used by automatic methods (\eg, recent frame interpolation models \cite{wan2025wanopenadvancedlargescale, gao2025seedance10exploringboundaries, po2025longcontextstatespacevideoworld, kong2024hunyuanvideo}) to generate in-between frames, a concept we term as \emph{\conceptname{}}.
Sketched notations have historically been rooted in animators' practice where arrows, lines, brief labels, and other informal notations are leveraged to externalize their thoughts, feel the dynamics in movements, and communicate with others~\cite{Green:1989:Cognitive, ware2010visual, thomas1981illusion, alikhani:2018:arrowsVerbs, Williams:2009:AnimatorSurvivalKit}.
These fast, intuitive, and expressive notations afford rich meanings such as forces, pose dynamics, paths, timing, and stylistic choices.

Sketch-based interfaces have been widely explored for animation support within and outside the HCI community for both animation authoring and modeling.
In the HCI community, interactive tools such as Draco~\cite{Kazi:2014:Draco}, Kitty~\cite{Kazi:2014:Kitty}, Energy Brush~\cite{Xing:2016:Energy-Brush}, Motion Doodle~\cite{thorne2004motion} map sketched notations to specific animation effects, facilitating the creation of kinetic textures, multi-subject causal illustrations, elemental dynamics, and character motions.
However, these tools rely on \emph{system-defined visual abstractions}, where the appearance and semantics of notations are strictly pre-determined. 
This enforces a fixed vocabulary that users must memorize, limiting the tools' generalizability to broader, improvisational animation scenarios.
Computer graphics and vision communities explored modeling approaches~\cite{Tanveer:2024:MotionBridge, Zhong:2025:Sketch2Anim} to treat sketched strokes as \emph{explicit} and \emph{exact} constraints that deterministically drive trajectories or subject movements.
However, in practice, animator's notations are often ambiguous and contextual: a line can signal arriving \emph{near} a position, emphasize timing or weight, or invite a humorous detour rather than prescribe a pixel-level accurate path.
Moreover, both lines of research usually handle one notation type at a time, overlooking how animators \emph{combine} notations to specify multiple aspects of animation effects.

In \emph{\conceptname{}}, we define sketched notations as \emph{user-defined visual abstractions} which are \emph{contextual}, \emph{ambiguous}, and \emph{combinational} guides for animation rather than pre-defined, strict, and isolated constraints.
By aligning more closely with artists' natural use of marks and symbols, we shrink the \emph{gulf of envisioning}~\cite{subramonyam2024bridging}, the gap between a creator's mental image and its externalization. 
However, the inherent ambiguity in notations shifts the burden to the system to interpret the intent, stretching the \emph{gulf of execution}~\cite{hutchins1986direct}; and it can also widen the \emph{gulf of evaluation}, as users must understand how the system interpreted their notations to iteratively refine the output. 
Thus, beyond supporting user-defined notation, a \conceptname{} system should also (i) provide a formalized interpretation layer to narrow the \emph{gulf of execution}, and (ii) establish a tight feedback loop to narrow the \emph{gulf of evaluation}.

As the initial step towards \conceptname{}, we first analyzed 135 real-world animator-produced sketches, in consultation with two professional animators (with over 25 and 15 years industry experience in animation respectively), and we categorized the meanings encoded in animators' notational marks.
From these observations, we propose a structured animation representation that formalizes the often informal notations for systematic interpretation. 
We then built a system probe with a two-level feedback mechanism to communicate the interpretation to users for quick confirmation and correction. 
Additionally, the system also exposes the lower-level parameterized control for motion range and timing through dynamically generated UI widgets to complement the high-level notations.
To test \conceptname{}, we conducted a preliminary user study with 7 professional animators to evaluate its usability, understand its affordance, and examine how it shapes animators' thinking and design workflows.
Our findings reveal that \conceptname{} enables a more holistic approach to keyframe authoring, shifting mental models from \emph{doing one thing at a time} towards \emph{thinking and composing multiple elements together}. 
We also identified practical constraints, most notably AI latency and performance, that raised model training considerations for the broader AI community. 
Finally, we offer design implications for exploring alternative animation representations across different abstraction levels, informing future research on AI-supported animation authoring.

In summary, we articulate the vision of \conceptname{} in this paper, along with the following contributions as the initial step towards realizing this vision: 
\begin{itemize}[leftmargin=*]
    \item The characterization of animators' notational practices;
    \item A formal animation intention representation maps the informal notations into system-executable actions; 
    \item A \conceptname{} prototype featured a two-level feedback mechanism and dynamic UI widgets for fine-grained control;
    \item The insights from a preliminary expert study detailing how professionals perceive and use \conceptname{}.
\end{itemize}

\section{Related Work}
\label{sec:related-work}

\subsection{Interactive Tools for Animation Authoring}

\subsubsection{Sketch-based Tools}

In HCI literature, there is a rich history of easy-to-use animation tools using sketching and direct manipulation that aims to make animation accessible to broader audiences. 
However, these tools predominantly rely on a \textbf{system-defined} approach: they pre-determine the mapping between visual abstractions and their semantics, requiring users to learn precisely \emph{what} they can sketch and \emph{how} to sketch within a rigid and procedural workflow which often complemented with UI widgets for parameter specifications.
Prior work has primarily explored two types of abstractions: those addressing \emph{space-time interactions} and those defining specific \emph{animation effects}.

\textbf{Space-time abstractions.}
One line of work addresses space-time interactions by mapping temporal relationships onto trajectory-based abstractions. 
K-Sketch~\cite{davis2008k} allows users to sketch object paths but relies on disparate UI widgets to specify transformations. 
Similarly, DirectPaint~\cite{Santosa:2013:Direct-Space-Time} adopts trajectory sketching for video annotation, using the path to propagate temporal changes. 
Some research~\cite{thorne2004motion, guay2015space} extends this space-time abstraction into 3D character animations.
These systems operate on a rigid keyframe-based logic where the stroke serves a single, fixed function (defining the path), while other properties, such as shape deformation, are managed via explicit parametric controls or algorithmic optimizations.

\textbf{Animation effect abstractions.}
Other systems represent \emph{specific} animation effects as distinct visual or gestural abstractions. 
Early work by Igarashi~\etal~\cite{igarashi2005rigid} mapped multi-touch gestures directly to shape deformation. 
Later, Motion Amplifiers~\cite{Kazi:2016:Motion-Amplifiers} reified animation principles (\eg, stretch and squash) as individual first-class visual objects which can be applied to drawings for exaggerated animation effects; and Energy Brush~\cite{Xing:2016:Energy-Brush} utilized sketched arrow as the abstraction to depict secondary dynamics like fluid flow of smoke, fire, water and other particles.
Such rigid mapping is also evident in Draco~\cite{Kazi:2014:Draco} and Kitty~\cite{Kazi:2014:Kitty}. 
Draco mapped kinetic texture effects to strokes, requiring users to explicitly classify their input stroke as either \emph{emitting} or \emph{oscillating}. This taxonomy confines the output to the system's predefined execution logic for translating strokes into animation.
Kitty abstracts these entities into a graph view, enabling users to further specify the functional relationships through sketching and direct manipulations.
Particular to character postures, prior work~\cite{Guay:2013:LineOfAction} adopted the line abstraction to quickly block out the shape of postures~\cite{Guay:2013:LineOfAction}.
The recent work Squidgets~\cite{kim2024squidgets} is perhaps the closest to our work, expanding stroke abstractions to support versatile effects like shape deformation, character rigging, and motion poses. 
However, this expansion introduces significant ambiguity. 
To manage this, Squidgets requires \emph{explicit} mode-switching: users must \emph{pre-select} a category (\eg, abstraction curves \vs~ rig curves) before drawing. 
This approach constrains users to the system's functional affordances and inhibits the combination of effects. 
Furthermore, ambiguity still persists even within categories (\eg, a shape stroke could imply scaling, moving, or deforming), yet the system offers only the \emph{best guess} without a feedback mechanism to resolve it.

While we draw inspiration from these prior works, they remain limited by \textbf{system-defined abstractions} that tailored to \emph{specific} animation effects with \emph{pre-determined} meanings that users must memorize. 
In contrast, we aim to provide an expressive \textbf{user-defined abstraction system} that leverages generative technologies to \emph{infer} meaning from \emph{generalizable} animation contexts, rather than enforcing rigid mappings, and to design a transparent feedback mechanism to resolve ambiguity dynamically.

\subsubsection{Generative Tools}
The emergence of generative models has led to new approaches in animation. Consumer tools such as RunwayML, Google Veo, and Adobe Firefly are capable of generating videos and animations from static images (functioning as keyframes) and/or video prompts. Although these methods are powerful and can produce high-quality results, they offer limited fine-grained control for stylized and expressive animations.

To facilitate control, conditional motion synthesis techniques ~\cite{Zhong:2025:Sketch2Anim, Wu:2025:DoodleYourMotion} generate detailed 3D character animations from 2D sketches. 
Alternatively, some reserch~\cite{Smith:2023:AnimatedChildrenDrawing} offers limited control by providing a library of pre-made animation templates, which users can select to automatically retarget motion onto their static drawings.
Generative video in-betweening frameworks ~\cite{Tanveer:2024:MotionBridge, wan2025wanopenadvancedlargescale, gao2025seedance10exploringboundaries, po2025longcontextstatespacevideoworld} offer versatile controls that encompass keyframes, trajectories, and masks to guide the generated frames. However, their outputs are pixel-level representations, which present challenges for editing and iteration. 
Moreover, the input trajectories function as rigid constraints, in contrast to our proposed notational animation inputs that accommodate greater abstraction, flexibility, and ambiguity. Although these methods show potential for creating detailed in-between frames, our work prioritizes the generation and manipulation of keyframes, which then serve as the foundation for these techniques.

In the field of HCI, researchers have explored the application of Large Language Models (LLMs) for generation and verification ~\cite{Ma:2025:MoVer} of animations from user prompts. 
This is achieved by generating animation code, rather than pixels, which allows for subsequent manipulation through code editing ~\cite{tseng2024keyframer}, keyframe editing ~\cite{Zhang:2023:Motion-Verification}, or dynamically synthesized UI widgets ~\cite{Liu:2025:LogoMotion}. 
More recently, Narrative Motion Blocks ~\cite{Bourgault:2025:Narrative-Motion-Block} demonstrates how direct manipulation input can be augmented with textual prompts to effectively create and edit animations. 
These recent works have shown the promising opportunity offered by generative models to shift from low-level parameter tuning to high-level intent specification. 
Yet, text is often ill-suited for capturing precise animation nuances. 
Although hybrid approaches like Narrative Motion Blocks~\cite{Bourgault:2025:Narrative-Motion-Block} combine text with direct manipulation, it is still an open question to determine the most effective interaction modality to communicate with generative models.
We argue that sketch-based input, with its well-established benefits in HCI, offers an intuitive and effective communication channel. 
However, a critical barrier, lies in shifting from the rigid, system-defined mappings of prior tools to a paradigm of \emph{user-defined abstraction} that supports \emph{generic} animation rather than specific effects. 
Our work addresses this gap by introducing a notational animation framework where users express intent primarily through sketched notations, supplemented by direct manipulation when precise parametric control is required.

\subsection{Expressing Intent beyond Natural Language}
\label{sec:related-work:intent}
Natural language alone is often insufficient to effectively convey intentions to AI~\cite{weisz2024design, chen2023next, simkute2024ironies, tankelevitch2024metacognitive}; Subramonyam~\etal~ terms this the \emph{gulf of envisioning}~\cite{subramonyam2024bridging}.
This reflects an \emph{instruction gap}: generative models are highly sensitive to phrasing, while people routinely use varied expressions to convey the same meaning.

A large body of recent work replaces or complements textual prompts with sketches, marks, and gestures to clarify intention. 
TaleBrush~\cite{chung2022talebrush} uses sketch lines to indicate narrative transitions, PromptPaint~\cite{chung2023promptpaint} provides a paint-like interface for semantic prompt interpolation,
and Block-and-Detail~\cite{Sarukkai:2024:Block-and-Detail} guides image generation with sparse strokes.
In animation production, Kaur~\etal~\cite{Kaur:2025:CHC} introduces a technique to overdraw on shots to communicate detailed hair and cloth motion for consistency across artists.
Gesture-based systems express edit intent with masks for inpainting \cite{avrahami2023blended}, colored strokes for recoloring \cite{zhang2017real}, and point dragging to adjust pose or facial expressions \cite{pan2023drag}. 
In programming, sketch-based interfaces have proven promising: Code Shaping~\cite{Yen:2025:Code-Shaping} supports the sketch-based expression of program intent, while Notational Programming~\cite{Arawjo:2022:NotationalProgramming} bridges handwritten and typewritten notations.

Another growing line of work augments natural language with direct manipulation by reifying user intent as manipulable GUI elements: 
DirectGPT~\cite{Masson:2024:DirectGPT} lets users drag visual elements into prompts, DynaVis~\cite{vaithilingam2024dynavis} integrates NLIs with dynamically generated widgets for visualization authoring. 
Design tools like Brickify~\cite{Shi:2025:Brickify} and AI Instruments~\cite{Riche:2025:AI-Instruments} reify intent into manipulable tokens or instruments for image composition and editing, extending Instrumental Interaction to AI contexts~\cite{beaudouin2000instrumental}.
Piet~\cite{shi2024piet} enables direct manipulation for color authoring in motion graphic videos.

In this work, we aim to support intent expression in animation authoring by using high-level sketched notations as the primary modality, complemented by dynamically generated UI widgets to balance simplicity and fine-grained control. 
Unlike prior work, we introduce an intermediate intent layer that formalizes user intent without constraining how notations are used.

\section{Background and Design Study}
\label{sec:design_space}

In \conceptname{}, we aim to leverage sketched notations as \emph{user-defined abstractions}, empowering users to employ their own visual vocabulary for animation keyframing without \emph{system-defined} constraints.
In this section, we first provide necessary background knowledge and define the scope of this work.
Then, we detail our methodology, covering both the content analysis and the expert interviews.
After that, we present the key observations and findings from the design study.
Finally, we distill the insights that inform our design of \conceptname{}.

\subsection{Scope and Terminology}

Although sketched notations appear in both storyboards and comics which have been extensively studied~\cite{mccloud1993understanding, mccloud2006making, eisner2008comics, abel2008drawing, cohn:2013:visual_language_of_comics, cohn:2015:notion_of_motion, cohn2020visual}, animation keyframing is fundamentally different from storyboarding and making comics. 
Following the definition in established animation book~\cite{Williams:2009:AnimatorSurvivalKit}, we clarify the scope of our work by distinguishing animation \emph{keyframing} from \emph{storyboarding and comics}, and then explaining the terminology \emph{keyframes} and \emph{in-betweens} in animation practices.

\textbf{Keyframing \vs~ Storyboarding.}
Storyboarding belongs to the \emph{planning} stage of animation. 
It operates at the scene or semantic level, defining narrative flow, staging, and camera choices. 
Each storyboard panel represents a discrete story beat (\eg, \emph{``the character enters the room''}). 
The space between panels implies \emph{semantic} jumps that must be mentally filled by the viewer, for example, changes in action, position, or perspective.
Keyframing, by contrast, belongs to the \emph{production} stage and operates at the motion or parameter level. 
A keyframe specifies the movement property of an object (\eg, position, rotation, pose) at a precise time. 
The gap between keyframes is not a narrative leap but a continuous, computable interpolation of movement. 
Keyframing begins only \emph{after} the storyboard has established the story and major poses.

\textbf{Keyframing \vs~ Making Comics.}
Comics are a distinct visual medium that aligns more closely with storyboards than with animation. 
Comic artists primarily convey \emph{what happens} (\ie, narrative beats), rather than \emph{how motion unfolds}. 
For example, depicting a character running in a comic requires only a single image; the full cycle of weight shifts and stride variations is unnecessary. 
In animation, however, such nuances are essential: the cadence of a run is important to shape the character's personality such as the iconic gait of \emph{Donald Duck}. 
These fine-grained motion details in keyframing are typically omitted in comics.

In summary, storyboards and comics share many similarities and aim to establish narrative structure, whereas keyframing focuses on designing the precise mechanics and style of motion. 
Consequently, prior studies~\cite{mccloud1993understanding, mccloud2006making, eisner2008comics, abel2008drawing, cohn:2013:visual_language_of_comics, cohn:2015:notion_of_motion, cohn2020visual} on notations in storyboards and comics cannot be directly generalized to keyframing practices, which we will discuss in Section~\ref{sec:notation_analysis}.

\textbf{Keyframes \vs~ In-betweens.}
Grounded in professional practice~\cite{Williams:2009:AnimatorSurvivalKit},
\emph{keyframes} refers to key poses that make an animation believable, which could be classified into three types: (i) \textit{keys}: the essential storytelling poses required for the action to read; (ii) \textit{extremes}: the turning points (\eg, contacts, anticipations) that define the limits of the movement; and (iii) \textit{breakdowns}: intermediate poses that determine \emph{how} the action travels (\eg, arcs, spacing) to ensure believability.
The typically workflow is iterative: animators typically draw \emph{keys} first, then iteratively add \emph{extremes} to structure the transition and \emph{breakdowns} to shape the motion's feel.
In this work, we collectively refer to all three categories (\ie, keys, extremes, breakdowns) as \textbf{keyframes}, which require artistic decisions from animators.
In contrast, \textbf{in-betweens} refer to all remaining transitional frames to make the motion fluid, which can be automated by computational interpolation methods.
In \conceptname{}, users draw the \emph{keys} as base drawings and iteratively add notations onto them to generate \emph{extremes} and \emph{breakdowns}. 
The resulting set of \emph{keyframes} is then interpolated by computational methods (\eg, a video model) to synthesize the \emph{in-betweens} for the final animation.

\subsection{Design Study Methodology}
\label{sec:content_analysis_method}

\begin{figure*}[htp!]
  \centering
  \includegraphics[width=\linewidth]{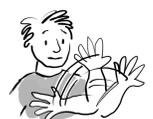}
  \caption{Three roles for animation notations emerged from the content analysis: geometric guides, spatial guides, and motion amplifiers. All visuals are redrawn by the authors to preserve IP while faithfully reflecting the originals.}
  \Description{ The figure summarizes three roles of animation notations with example sketches: Geometric guides (hidden from the audience, and the paper’s focus) include marks for force (arrows showing cause/direction/magnitude), path (curves, dotted trails, arrows for trajectories and secondary motion), pose (S/C ``line of action'' capturing overall dynamism), and style (mark qualities like sharpness, jitter, thickness conveying motion feel); Spatial guides (also hidden) include arcs for smooth alignment across frames and perspective guides for depth/scale changes; Motion amplifiers (visible to the audience) include motion lines, impact stars, vibration tics, blur/streak substitutions, background streaks suggesting camera follow, and repeated body parts to depict movement over time.}
  \label{fig:notation_types}
\end{figure*}

Our design is informed by domain insights obtained from three sources: (i) established practices illustrated in the influential book in animation \emph{The Animator’s Survival Kit}~\cite{Williams:2009:AnimatorSurvivalKit}; (ii) a content analysis of animation notations across 135 real-world sketches; and (iii) interviews with two professional animators with 25 and 15 years of industry experience respectively (noted as E1 and E2) to validate and contextualize our observations. 

\subsubsection{Data Collection}
We collected a corpus of notation-embedded animation sketches (135 in total) from the online platform \emph{Pinterest} where animators share work and foundational books~\cite{Williams:2009:AnimatorSurvivalKit, Gilland:2012:ElementalMagic}. 
We searched using the key terms including \emph{``animation sketch''}, \emph{``motion sketch''}, \emph{``animation notation''}, \emph{``animation notes''}, \emph{``pose notation''}, \emph{``animation action sketching''}.
We selected the sketches that contain at least one graphical notation (\eg, arrows, lines) and excluded examples containing only textual descriptions. 
To capture diversity across contexts, we balanced entities such as human characters, animals, cartoon characters, objects, and particle systems (\ie, water, fire, cloud, etc.). 
To reduce redundancy, we removed near-duplicates; for instance, among many sketches depicting walking in similar notating ways, we only keep the most representative one.
The resultant corpus of the 135 examples can be found in the Appendix~\ref{sec:coding_result}.
Due to copyright constraints, we provide external links for all examples rather than displaying the images directly.

\subsubsection{Coding}
Prior to coding, we reviewed the book~\cite{Williams:2009:AnimatorSurvivalKit} telling the tacit knowledge in animation to ground our domain understanding and the nuances animators consider. 
Informed by this, the first author conducted inductive open coding of each example, annotating all notations and their intended meanings. 
Then, we employed an iterative procedure until no new codes emerged.
After each round, the co-authors reviewed code meanings and groupings together; the codebook was refined through discussion. 
Following the iterations, the first author applied the codebook in a deductive recoding of the full corpus to ensure consistency.
After that, we interviewed two professional animators (E1, E2) to examine the corpus, code definitions, and groupings, suggest clarifications, and confirm that no obvious notation types were missing.
We incorporated their feedback and completed a final re-coding pass. 
We include the detailed codebook in Appendix~\ref{sec:coding_scheme} and the full coding results in Appendix~\ref{sec:coding_result}. 
The outcome of this analysis defines a broad framework and categories of animation notations.
\subsection{Observed Notation Types and Meanings}
\label{sec:notation_analysis}

Within the defined scope of animation keyframing process, we now consider the notational marks that often leveraged in keyframe drawings. 
In practice, animators augment drawings with quick annotations to externalize ideas and communicate with collaborators. 
As E2 noted, \qt{I love to use it to quickly capture my thoughts ... this way is intuitive and fast, letting me plan the whole and mingle decisions across parts.} 
However, these notations are diverse and context-dependent. 
E1 emphasized that \qt{this kind of language (notation) varies across studios; conventions could differ.} 

To examine these practices, we conducted a content analysis of 135 animator-produced keyframe sketches (Section~\ref{sec:content_analysis_method}) and distilled recurring patterns in how motion and intent are marked. 
In specific, we categorized them into three groups: \textit{geometric guides}, \textit{spatial guides}, and \textit{motion amplifiers}.
It should be noted that we do not intend to propose formal definitions or definitive categories here.
Our aim is to depict the scope and highlight identified recurring patterns (Figure~\ref{fig:notation_types}) to support a shared understanding for animation notations.
We describe the three identified categories below; detailed descriptions of each sub-category are embedded in Figure~\ref{fig:notation_types} alongside representative examples from our corpus. 
To avoid intellectual property concerns, all visuals were redrawn by the authors, with notations that faithfully reproduce the originals.
The full corpus with the coding results can be found in Appendix~\ref{sec:coding_result}.

\subsubsection{Geometric Guides.}
Geometric guides are notations that specify how an object’s shape should deform or a subject's kinematics should change to realize an animation.
This is the most common type of notation we found in our corpus (with 102 instances). 
We categorize these guides into four main purposes: force, path, pose, and style (the most representative examples for each category are shown in Figure~\ref{fig:notation_types}). 
For each category, we further code them into detailed meanings, the detailed coding scheme can be found in Appendix~\ref{sec:coding_scheme}.
These notations are planning aids for animators and will be removed before the final render; audiences never see them.
We summarized the following key characteristics (noted as C1-4).

\textit{\textbf{C1}: The semantics of notation are highly \textbf{contextual}.} 
Across the four categories (\ie, force, path, pose, and style), animators most often use arrows, lines, and curves, and occasionally other shapes (\eg, circles, rectangles, triangles).
Yet these marks have no fixed \emph{form – meaning} mapping: the same mark (\eg, an arrow) can represent different concepts across categories. 
Text, numbers, or color are sometimes added to clarify motion or timing, but these cues are often omitted and interpreted contextually (\eg, numbers may indicate sequence or a region of interest). 
In summary, unlike linguistic systems constructed by rigid syntax and grammar, animators' notations are flexible and derive meaning from context.

\begin{figure}[htbp]
    \includegraphics[width=\linewidth]{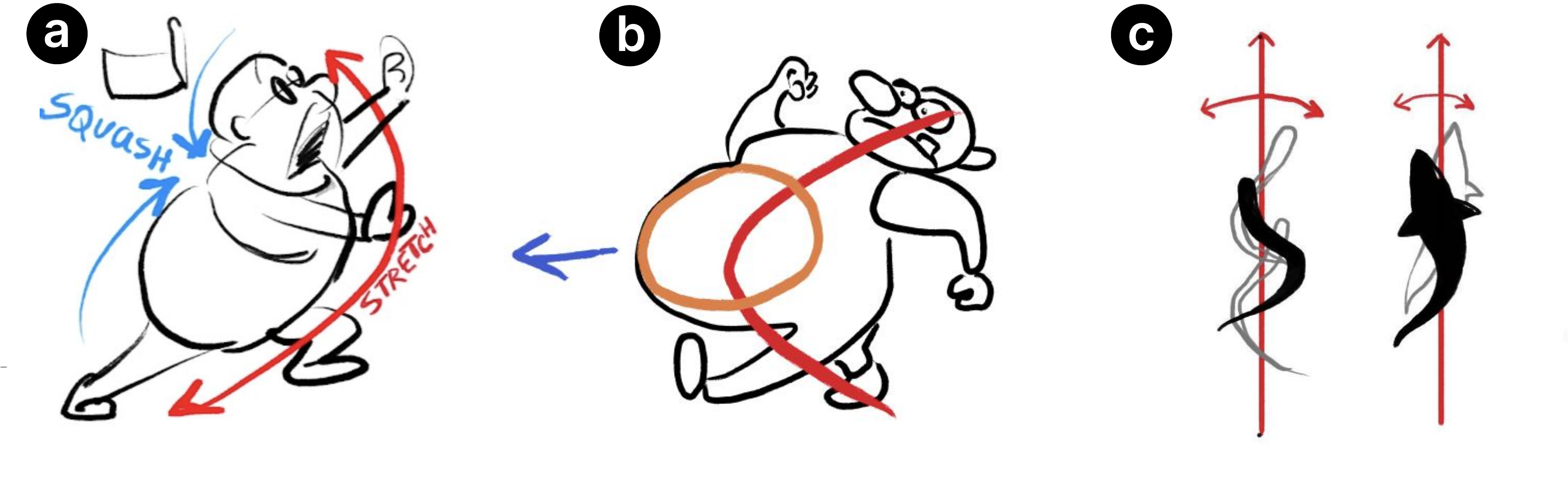}
    \vspace{-8mm}
    \caption{Combinational usage of notations: (a) combining notations all related to \emph{force}; (b) combining notations of \emph{pose} and \emph{force}; (c) combining notations for \emph{primary motion} and \emph{secondary motion}.}
    \Description{Three sketches labeled (a–c) show how different animation notations can be combined. (a) A squashed character pose is overlaid with curved arrows and ``squash/stretch'' marks to indicate applied force. (b) A running character is annotated with a large looping/spiral stroke and a direction arrow, combining pose with force cues. (c) Two vertical guides with rotation arrows and wavy trails contrast primary motion (main twist/turn) with secondary motion (additional trailing wobble).}
    \label{fig:combinational_notations}
\end{figure}

\textit{\textbf{C2}: The usage of notations is often \textbf{combinational}.} First, multiple notations conveying the \emph{same} category of meaning (\eg, all related to force) can appear in a single drawing to depict individual motions for different parts. 
For example, Figure~\ref{fig:combinational_notations}.a shows how the body stretches and squashes at the same time. 
Second, notations representing \emph{different} categories of meaning (\eg, force \vs~ pose) can be combined to depict multiple aspects of an animation such as Figure~\ref{fig:combinational_notations}.b.
Third, multiple notations can be used in combination to serve \emph{hierarchical} roles, such as the primary motion in combined with secondary motions, as shown in Figure~\ref{fig:combinational_notations}.c.


\textit{\textbf{C3}: The meanings of notations are inherently \textbf{ambiguous}.} 
This ambiguity manifests in three primary aspects. 
First, regarding \textbf{scope}, notations are often placed near a specific body part to indicate where the animation applies without explicitly specifying the target object. 
Thus, the affected area can only be inferred from context: it could refer to the entire character, a local limb, or even multiple connected parts. 
Second, regarding \textbf{spatial precision}, notations are expressive in describing high-level intent about what the motion is and how it feels, but rarely specify \emph{pixel-level} accuracy. 
For example, an upward arrow indicating a jump with its rough intensity but does not necessarily mean the exact height which instead \pqt{could be iteractively tuned to find the right one (magnitude).}{E2}
Lastly, ambiguity arises in \textbf{timing}. 
Numerical annotations can indicate action order, but animation timing is more complex, requiring coordination of durations and overlaps.  
Such temporal ambiguity likely because static spatial marks are short in conveying duration or speed. 
As noted by E1, these details remain abstract: \qt{they are acting in my head but I won't write it down.}

\subsubsection{Spatial Guides.}
Unlike geometric guides that mostly annotate a single frame, spatial guides often span multiple frames. 
They map how positions and scales should change over time and align arcs, paths, and sizes across frames with lines or curves. 
Like geometric guides, they are temporary construction aids and are removed before the final animation.
There are two primary categories in spatial guides: proportion alignments and perspective guides. 
We explain them in details with representative examples in Figure~\ref{fig:notation_types}.

\subsubsection{Motion Amplifiers.}
As this category overlaps significantly with prior studies in comics~\cite{mccloud1993understanding, mccloud2006making, eisner2008comics, abel2008drawing, cohn:2013:visual_language_of_comics, cohn:2015:notion_of_motion, cohn2020visual}, we adopt their established taxonomy and show the examples for the six different kinds of amplifiers in Figure~\ref{fig:notation_types} but omit a detailed review here.
The key difference distinguishing motion amplifiers from geometric and spatial guides is their function: amplifiers help the \emph{audience} feel the motion effect in a static medium, rather than assisting the \emph{creator} in thinking about the motion. 
Thus, notations for geometric and spatial guides are never visible to the audience, whereas motion amplifiers are deliberately created for audience.

\subsubsection{Focusing on Geometric Guides.}

In this paper, we narrow down our focus to support \emph{geometric guides}. 
We exclude \emph{motion amplifiers} because of two reasons. 
First, they function primarily as \emph{audience-facing} perceptual cues, helping viewers perceive motion in static media like comics, rather than as \emph{creator-facing} tools for thinking and planning motion dynamics. 
This distinction was confirmed by experts E1 and E2. 
Second, even if \emph{motion amplifiers} could be re-purposed for authoring, integrating them will result in a \emph{system-defined notational system} where users have to learn and select from a fixed library of symbols.
It is because they are well-established visual conventions in comics with deterministic meanings for each.
This conflicts with our goal in this work to enable \emph{user-defined notations} that users can flexibly define their own notations in free-form.
We thus limit our focus to notations that stay informal and externalize animator's intent directly.

Additionally, while \emph{geometric} and \emph{spatial guides} assist in motion planning, they serve distinct roles: \emph{geometric guides} operate \emph{locally} to define the internal structure of individual keyframes, whereas \emph{spatial guides} operate \emph{globally} to define composition across multiple keyframes.
We argue these processes are functionally decoupled and present different interaction needs. 
Spatial alignment can be intuitively handled via direct manipulation (\eg, move, scale, and rotate existing keyframes). 
In contrast, crafting the geometric deformations of animated subjects, such as designing their poses, requires much creative efforts.
While AI could theoretically assist in spatial alignment as well, the current bottleneck for animators remains in the \emph{generation} of these individual frames, not their \emph{placements}. 
Therefore, to avoid over-complicating the problem space, we prioritize the most critical problem of generating individual keyframes that align with the animator's intent expressed through \emph{geometric guides}. 
We reserve the integration of \emph{spatial guides} for future iterations, with an envisioned direct manipulation-based interaction approach similar to StickyLines~\cite{ciolfi2016beyond}.

\subsection{Design Goals}

Based on above observations from animators' sketches, with a focus on the \emph{geometric guides}, and our discussion with professional animators, we distilled the following design goals to guide the design of a \conceptname{} system.

\textbf{DG1: Support freeform notation without rigid syntax.} 
To accommodate the freeform nature of sketching, the system should allow users to notate flexibly (\textit{C1: contextual semantics}) without enforcing a fixed vocabulary or grammar. 
That is, how to notate should be decided by user themselves instead of the system.

\textbf{DG2: Accommodate combinational usage for multifaceted intentions.} 
As animation design decisions are inherently intertwined, users often express multiple intentions within a single drawing. 
The system must support arbitrary combinations of notations \textit{(C2: combination usage)}, enabling it to recognize and process multiple constraints (\eg, force and pose) simultaneously.

\textbf{DG3: Provide structured interpretation and transparent feedback.} Given the contextual dependency and inherent ambiguity of notations, using them to instruct an automated system requires transparency.
Thus, the system must establish a closed feedback loop that communicates the AI's interpretation to users, including \emph{what} the notations mean \textit{(C2: contextual semantics)} and \emph{how} they apply to the animated subjects \textit{(C3: ambiguous scope)}. 
The feedback should minimize cognitive load for at-a-glance verification, with intuitive mechanisms for quick error correction.

\textbf{DG4: Facilitate iterative tuning through complementary interaction modalities.} 
Since notations fall short in conveying specific spatial magnitudes and temporal durations (\textit{C3: ambiguous spatial/temporal precision}), the system should provide interaction modalities beyond notation sketching to allow users to iteratively adjust motion range and timing.

\section{\sysname{}}

Informed by the characteristics of animators' notations (C1-3) identified in our design study and the distilled design goals (DG1-4), we define the concept of \conceptname{} and articulate its principles. We then introduce a \emph{formalized animation representation} for systematically interpreting notations and describe key features in a \conceptname{} system. 

\subsection{Definition}

We propose \conceptname{}, an interaction paradigm for animation keyframing, in which users sketch \emph{high-level notations} over static drawings to indicate intended animations. 
With the reasoning capability, a Vision-Language Model (VLM) interprets these notations into structured intent, which is used by a generative model (\eg, an image generation model) to synthesize animation \emph{keyframes}.
The resulting \emph{keyframes} can then drive computational interpolation approaches, such as recent video models, to produce \emph{in-between} frames for the final animation. 
In the following, we introduce the \textbf{scope} of \conceptname{}, the \textbf{role} of notations, and the \textbf{interaction modalities} in a \conceptname{} system.

\begin{itemize}[leftmargin=*]
    \item \textbf{Scope: a generalizable interaction paradigm.}
    We intend to design \conceptname{} for broad and general animation authoring instead of specific animation effects, including character animation, particle system animation, logo animation, etc.
    Thus, when interpreting the notations, any domain-specific assumptions should not be made. 
    For example, it should not assume a set of possible animations and select the most closed one for a notation, but all meanings should be inferred from the context.
    
    \item \textbf{Role: notations as user-defined abstractions.}
    The notions in a \conceptname{} system, serving as \emph{user-defined abstractions} to express intended animation,
    should stay \emph{informal}, in contrast to the \emph{formal notations} in math, physics, and music where a universal convention should be learned before using. 
    The notations in a \conceptname{} should also be \emph{contextual}, \emph{ambiguous}, and \emph{combinational} as discussed in Section~\ref{sec:design_space}.

    \item \textbf{Interaction: sketching combined with direct manipulation.}
    In a sense, \conceptname{} does not mean that users should only rely on sketching notations. We recognize that notations cannot afford all intentions. 
    Thus, in a \conceptname{} system, to fully realize artists' vision, notations should be used together with direct manipulation. Notations are for high-level expression and direct manipulation is for low-level control, especially on the motion spatial range, effect intensity, and timing coordination. 
\end{itemize}

\subsection{Structured Animation Representation}
\label{sec:animation_representation}

Instead of formalizing notations, we choose to formalize the intended animation representation to facilitate the interpretation of the notations (\textit{DG1: support freeform notation}).

\vspace{2mm}
\begin{aptdispbox}
\acmcallout{Animation Notation Representation:}{
\label{box: source-path-target}
\textbf{Primary Notations:}  <{\it Source}, {\it Path}, {\it Target}> to express animation semantics:
\begin{itemize}\leftskip5pt
    \item[\textit{Source}:] where motion originates, what components involved to move, could be implicitly indicated by proximity or explicitly marked by users.
    \item[\textit{Path}:] how motion unfolds, it could be [force | trajectory | direction | etc.], while often shown by arrows, it can also be conveyed through lines without arrowheads.
    \item[\textit{Target}:] how the intended end status looks like or where the motion ends. The line of action could serve as one kind of ``target'', but other forms are possible, such as a simple positional mark indicating where the source should move to, or the drawing of a detailed pose.
\end{itemize}
   
\textbf{Secondary Notations:} to express stylistic descriptors of an animation and add additional information or clarifying primary notations. [color | thickness | numeric labels | text labels | etc.] could be used to convey temporal order, motion style, intensity, or other expressive metadata.
}
\end{aptdispbox}
\vspace{2mm}

This representation is based on observations from our corpus (\textbf{C1: contextual semantics}).
We categorize notations into two types: primary notations and secondary notations, according to their appearing frequency and function.
Primary notations consist of frequent graphical notations (\eg, arrows, lines, curves) that define the core \emph{semantics} of the animation (\eg, force, pose, path). 
However, since notation semantics can be ambiguous regarding the concrete animation behavior and the targeted object (\textbf{C3: ambiguous scope}), we decompose the meaning of primary notations into a structured triplet: $\langle source, path, target \rangle$.
Secondary notations consist of auxiliary but often omitted elements (\eg, color-coding, text, numbers) that provide \emph{supplementary information}, such as temporal ordering or the expressive feeling of the motion.
We highlight three key considerations during the interpretation process:

\begin{figure}
    \centering
    \begin{subfigure}[t]{\linewidth}
        \includegraphics[width=\linewidth,trim={0 165em 0 0 },clip]{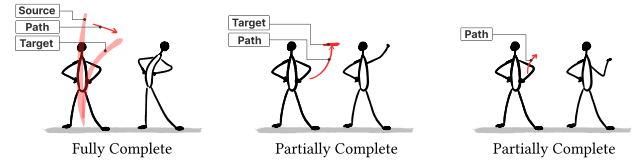}
        \vspace{-6mm}
        \caption{Illustration of the three core fields in our formalized notation—source, path, and target. Left to right, examples vary in how completely these fields are specified. When any field is omitted, the system infers the missing information from context.
        }
        \Description{Three side-by-side examples illustrate the notation's three core fields (\ie, source, path, and target) using a stick figure and red arrows. From left to right, the examples show complete primary notation (all three fields labeled), partially complete notation (one field omitted), and another partially complete case; missing fields are intended to be inferred from context.}
        \label{fig:flexible_completeness}
    \end{subfigure}

    \begin{subfigure}[t]{\linewidth}
        \includegraphics[width=\linewidth,trim={0 0 0 160em },clip]{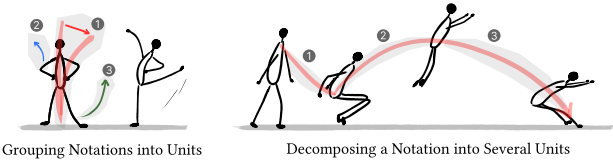}
        \caption{Illustration of grouping (left) and decomposing (right) notations into multiple animation units.}
        \Description{Two panels show how notations map to animation units. Left: several annotated gestures around a stick figure are grouped into one combined unit. Right: one long curved motion notation is decomposed into multiple units, illustrated by a sequence of stick-figure poses along a jump/arc trajectory.}
        \label{fig:units}
    \end{subfigure}

    \vspace{-3mm}
    \caption{Illustration of flexible completeness and the demonstration of multiple animation units in one sketch.}
    \label{fig:notation_representation}

\end{figure}

\begin{itemize}[leftmargin=*]
    \item \textbf{Visual Form Agnosticism:}
    This representation places \emph{no constraints on the visual form} of notations (whether it is an arrow, a line, or a circle).
    Instead, it prioritizes each notation's functional role (\ie, source, path, target) in the intended motion (\textbf{C1: contextual semantics}).
    Additionally, this representation supports compositional notations: rather than assuming a one-to-one mapping between a single mark and an animation, it recognizes that a single animation unit is often composed of multiple distinct notations (\textbf{C2: combinational usage}). 
    For example, a user may use a circle to specify the subject (source), an arrow to define the trajectory (path), and a line to indicate the final position (target).

    \item \textbf{Atomic Definition:}
    This representation \emph{declares a single animation unit}, though users may specify multiple units simultaneously (\textbf{C2: combinational usage}), as shown in Figure~\ref{fig:units}.
    Thus, during interpretation, notations should be first grouped into units, each unit is then interpreted with this representation (\textit{DG2: accommodate multifaceted intentions}). 
    On the other hand, a single notation may denote multiple units (\eg, in Figure~\ref{fig:units}, the leap consists of three units). 
    Thus, during interpretation, a notation may need to be decomposed into multiple animation units.

    \item \textbf{Flexible Specification Completeness:}
    This representation allows for \emph{flexible completeness}. 
    Users are not required to specify the full triplet $\langle source, path, target \rangle$, nor are secondary modifiers mandatory (\textbf{C3: ambiguous meaning}). 
    When all elements are present, the system interprets the motion intent with high precision. 
    Conversely, when elements are missing, the system accommodates \emph{creative ambiguity}; the interpreter (\eg, vision-language models) infers the missing fields from the context. 
\end{itemize}

\begin{figure*}[tp!]
  \centering
  \includegraphics[width=\linewidth]{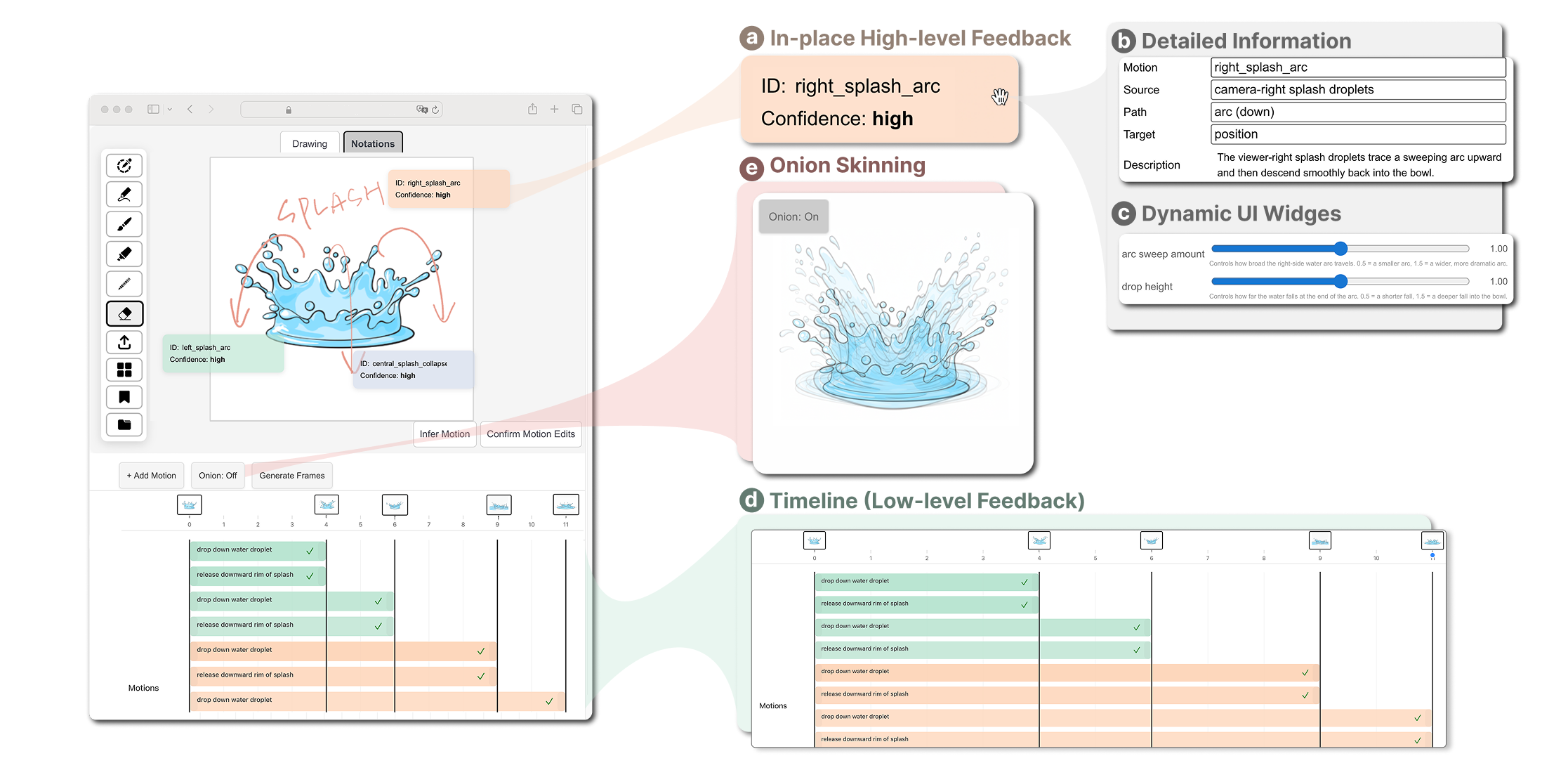}
  \caption{Interface of our \conceptname{} system: the main canvas is augmented with (a) motion tags for in-place high-level feedback, (b) hoverable detailed information, (c) dynamically generated sliders for motion range adjustments, (d) an auto-constructed timeline for timing edits and per-object feedback, and (e) onion-skin overlays to review keyframe movement.}
  \Description{A composite UI figure shows the notational animating system. On the left, a main drawing canvas contains a splash illustration annotated with handwritten motion marks and small motion-tag callouts (ID and confidence). On the right, zoomed panels highlight key features: (a) in-place high-level feedback tags, (b) a hoverable details table for a motion (source/path/target/description), (c) dynamically generated sliders to adjust motion ranges, (d) an auto-built timeline for timing edits and per-object feedback, and (e) an onion-skin overlay preview of keyframe movement.}
  \label{fig:interface}
\end{figure*}

This formalization serves two main purposes.
First, it resolves ambiguity by requiring the system to explicitly infer essential components from the context. 
Our experiments indicated that without such structural constraint, VLMs tended to default to high-level, global semantic descriptions. 
In contrast, enforcing this structure guides the model to produce fine-grained interpretations that align better with the animator's specific intent. 
We provide an example in Appendix~\ref{sec:vlm_responses} comparing VLM interpretations with and without this structural constraint.
Second, such a formal structure could allow for providing users with the \emph{structured and verifiable feedback}. 
Without an explicitly defined structure, interpreters return free-form text that varies unpredictably across different cases and models. 
This variability makes it difficult to visualize the AI's understanding effectively. 
Presenting users with lengthy, unstructured text imposes a significant cognitive burden, hindering their ability to quickly verify and correct the interpretation.

\section{The \sysname{} Prototype}

We implemented a prototype system as a testbed for \conceptname{}. 
Our prototype consists of three main components: (1) a drawing canvas to sketch and iteratively refine notations; (2) a two-level feedback mechanism that communicates how notations are interpreted and resolve ambiguities (\textit{DG3: provide structured feedback}); and (3) direct manipulation widgets (\ie, dynamic UI and timeline) for fine-grained control about how each individual subject parts move (\textit{DG4: facilitate iterative tuning}).

\begin{figure*}[htp!]
  \centering
   \includegraphics[width=\linewidth]{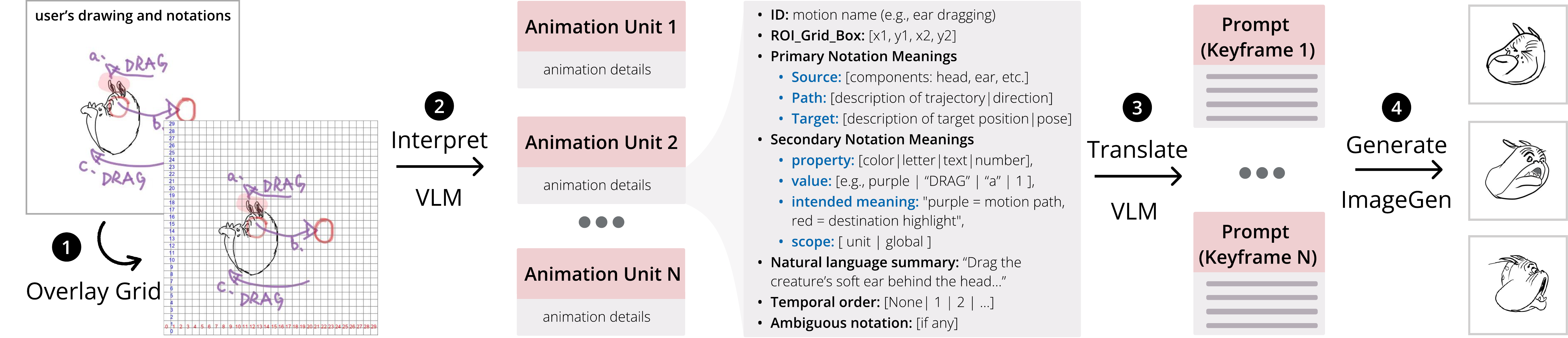}
  \caption{The technical pipeline of using VLM to interpret the sketched notations, consisting of four main steps: (1) overlay a grid onto user's drawing and notations; (2) VLM interpret it by identifying the animation units and interpret each unit in a structured way; (3) VLM translate the structured interpretation into natural language as the prompts for each keyframe; (4) parsing to an image generation model to generate keyframes.}
  \Description{A left-to-right pipeline diagram shows four steps for turning sketched animation notations into keyframes with a VLM: (1) overlay a grid on the user's drawing and drawn arrows/marks, (2) the VLM interprets the sketch by extracting multiple ``animation units'' with structured fields, (3) the VLM translates each unit into a text prompt for each keyframe, and (4) an image-generation model produces the resulting keyframe images.}
  \label{fig:vlm_pipeline}
\end{figure*}

\subsection{Drawing Canvas}
\label{sec:drawing canvas}
The canvas allows an animator to draw the static frame and notate on top of it by switching between the \textit{drawing} and \textit{notations} modes at the top of the canvas. 
Besides drawing from scratch, users can also upload an image. 
The drawing and notating brush's size and color can be changed by the tools provided in the toolbar at the left side of the canvas. 
The system treats the drawings and/or user-uploaded image in the same way as a rasterized image. 
The animator can save the current workspace (including all drawings, notations, status of the timeline, and results) by the \textit{Save Frames} button in the toolbar and revisit previous ones in \textit{History}. 

\subsection{In-place Motion Labels}
\label{sec:in-place feedback}
After the user finishes drawing and adding notations, clicking \textit{Infer Motion} sends the drawing or uploaded image together with notations to a vision–language model (VLM) for interpretation. 

\textbf{High-level feedback for animation interpretation} (\textit{DG3: structured feedback}).
Each animation unit is returned as a colored \textit{in-place motion label} (Figure~\ref{fig:interface}a), with colors randomly assigned per unit, positioned adjacent to its corresponding notation(s).
Each label shows a brief semantic description and a confidence rating (\ie, low, medium, or high) and can be dragged to reorganize the workspace. 
Hovering it will reveal details (Figure~\ref{fig:interface}b) about its inferred $\langle source, path, target \rangle$ and a brief description. 
We surface this information structure to give users an ambient mental model of how the system interprets their notations and to reason about how to communicate more effectively.
These details are editable, enabling users to correct any AI misinterpretations.
After making changes, the user clicks \textit{Confirm Motion Edits}; the system incorporates the edits and re-infers the motion accordingly.

\textbf{Dynamic UI widgets for fine-grained control.}
In parallel, the system generates context-aware slider widgets (Figure~\ref{fig:interface}c) for fine-grained control over spatial range and motion intensity (\textit{DG4: iterative tuning}). 
Default maximum and minimum values are provided. 
For quantitative parameters (\eg, rotation degrees), numeric values are shown; for qualitative ones (\eg, how high the hand raises), slider endpoints are labeled with semantic anchors (\eg, minimum = shoulder level).
It is worth noting that alternative design choices exist for providing fine-grained control over motion intensity and spatial ranges. 
Ideally, the most precise and intuitive method would be \emph{direct manipulation via on-canvas control handles}. 
However, integrating such interactions into current generative AI pipelines presents significant feasibility challenges.
First, since the system operates on raster images, generating the necessary semantic handles such as kinematic skeletons for characters requires non-trivial engineering efforts. 
Second, even if users were to manipulate pixels directly, current generative models struggle to adhere to such precise pixel-level guidance without extensive fine-tuning. 
Thus, we adopt dynamic UI widgets as a pragmatic solution to offer fine-grained control within current technical constraints. 
We envision that future advancements will eventually enable more direct and precise manipulation approaches.

\subsection{Timeline}
\label{sec:timeline feedback}
The timeline ((Figure~\ref{fig:interface}d) is automatically constructed once in-place feedback is generated and serves twofold purposes.

\textbf{Low-level feedback for how primitive parts move} (\textit{DG3: structured feedback}). 
For each \emph{animation unit}, the system decomposes motion into per-part tracks.
Each timeline block represents a primitive part (\eg, \textit{a character's head}) and is labeled as \emph{primitive's name} -~\emph{verb} (\eg, \emph{head tilt up}). 
Timeline block colors mirror the in-place feedback labels for easy mapping: all per-part timeline blocks belonging to the same animation unit share the color of its in-place \textit{motion label}.
By hovering over each block, a detailed description about how this part moves will be shown.

\textbf{Temporal order and duration adjustment.} 
The timeline allows for direct manipulation of temporal order and duration (\textit{DG4: iterative tuning}).
Blocks can be moved along the time axis, resized to adjust duration by dragging their edges, deleted, or added. 
By default, a keyframe is appended to each block, shown as black vertical lines with thumbnail previews above the timeline; hovering enlarges a preview, and clicking places the keyframe on the canvas.

When satisfied with the interpretation and timeline, the user clicks \textit{Generate Frames}; the system fills all empty keyframe placeholders with generated keyframes.
By toggling on the \emph{onion-skinning} (Figure~\ref{fig:interface}e) button, user-selected keyframes will be overlayed on the canvas to visualize the animation.

\begin{figure*}
    \centering
    \includegraphics[width=\linewidth]{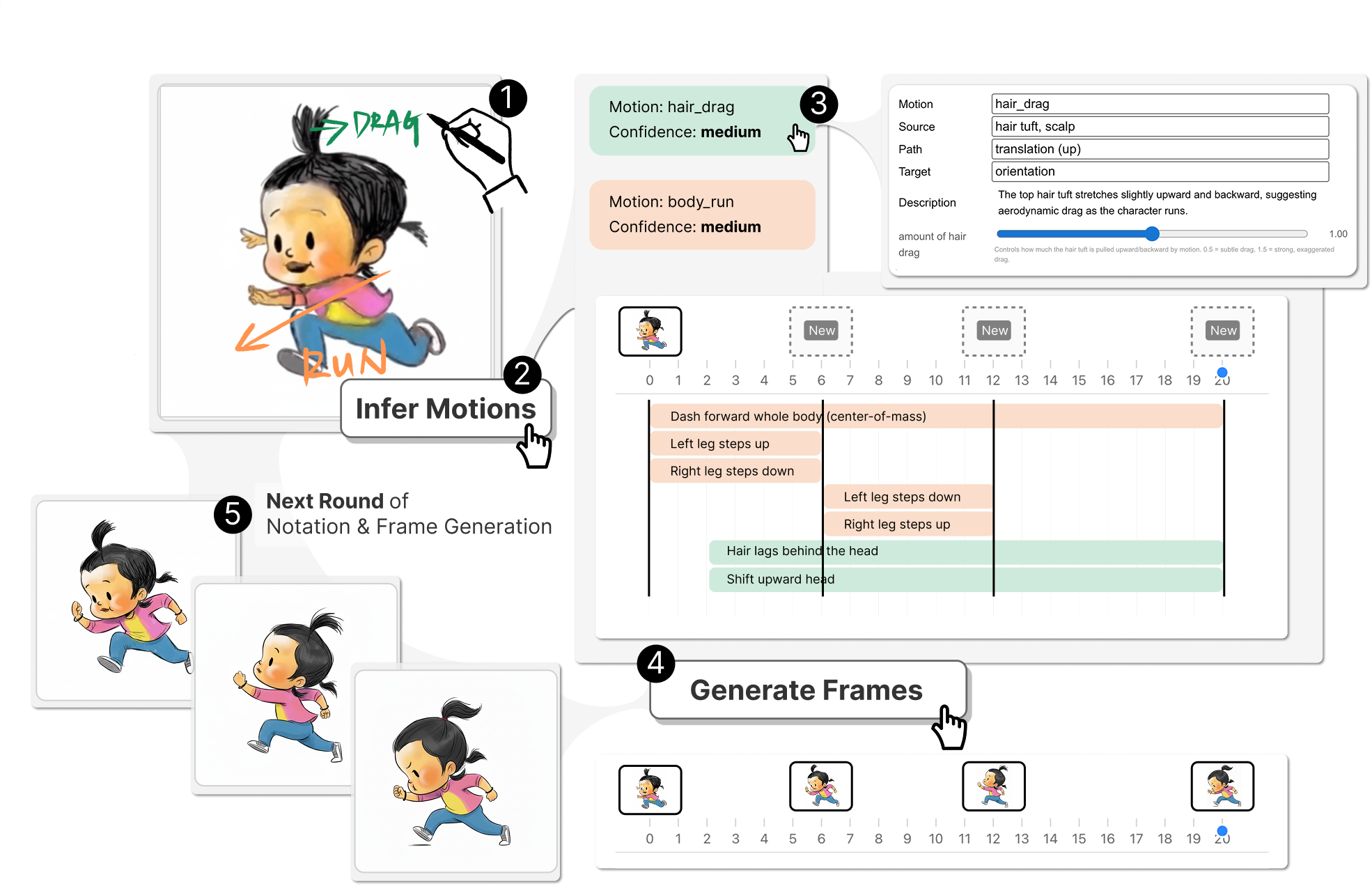}
    \caption{Workflow with \conceptname{} for character motion: (1) sketch high-level notations on the drawing; (2) click ``Infer Motion''; (3) the system overlays motion tags near the marks, hovering a tag reveals details, and automatically constructs a per-object timeline; (4) click ``Generate Frames'' to produce the next three keyframes; (5) continue notating on the generated frames to progressively author a longer keyframe sequence.}
    \Description{A step-by-step workflow diagram for notational animating a running character. (1) The user draws high-level marks on the illustration (e.g., ``drag'' on hair, ``run'' on body). (2) Clicking Infer Motions triggers the system to recognize motions. (3) Motion tags (e.g., hair_drag, body_run) appear with confidence and can be hovered to view detailed fields (source/path/target). (4) A per-object timeline is auto-generated and the user clicks Generate Frames to produce the next keyframes. (5) The generated frames are shown, and the user continues annotating to extend the sequence.}
    \label{fig:usage_example_01}
\end{figure*}

\subsection{Implementation Details}

Our system utilizes a Vision-Language Model (VLM) to interpret user notations into structured animation intentions and an image generation model to generate keyframes.
In specific, we use OpenAI’s \texttt{o3} API for (i) motion inference, (ii) decomposing each animation unit into primitive motions, and (iii) generating detailed keyframe descriptions. 
Keyframe images are produced with the \texttt{gemini-2.5-flash-image-preview} API. 
Our pipeline consists of four main stages (Figure~\ref{fig:vlm_pipeline}). 

\textbf{Motion interpretation.}
To enable precise localization, we first employ a \emph{spatial grounding} technique by overlaying a $30 \times 30$ coordinate grid on the input sketch. 
This grid provides a fixed reference system, enabling the VLM to ground observations in specific numerical coordinates rather than vague spatial descriptions.
Next, the VLM decomposes the sketch into ``animation units.'' 
We employ a structured prompting strategy (Appendix~\ref{sec:appendix}) to extract a formal representation (see Section~\ref{sec:animation_representation}) for each unit, returned as a JSON object containing unique identifiers and Region of Interest (ROI) coordinates for each unit.
This structured output enables the interface to synthesize the interpretation tags with semantic data and utilizes the ROI coordinates to spatially anchor these tags adjacent to their corresponding notations.

\textbf{Keyframe generation.}
Following interpretation, the VLM translates these structured specifications into natural language prompts optimized for image generation. 
By integrating inferred temporal logic with user-manipulated timeline constraints, the system synthesizes frame-specific prompts that detail both local component movements and global scene states.
Finally, these prompts and the initial drawing will be sent to an image generation model. 
For multi-frame sequences, we adopt a progressive strategy: each generation step is conditioned on the previously generated keyframe and the target prompt of the next, iterating frame-by-frame to ensure temporal continuity and smooth motion.

\section{Demonstrative Examples}

In this section, we present three illustrative examples produced with our \conceptname{} prototype. 
We provide a detailed account of the creation process for the first example to showcase the \conceptname{} workflow. 
For the second and third examples, we omit step-by-step details and instead highlight how the concept generalizes across diverse animation contexts and intentions.

\paragraph{Character Motion}

\begin{figure}
    \vspace{3mm}
    \includegraphics[width=\linewidth]{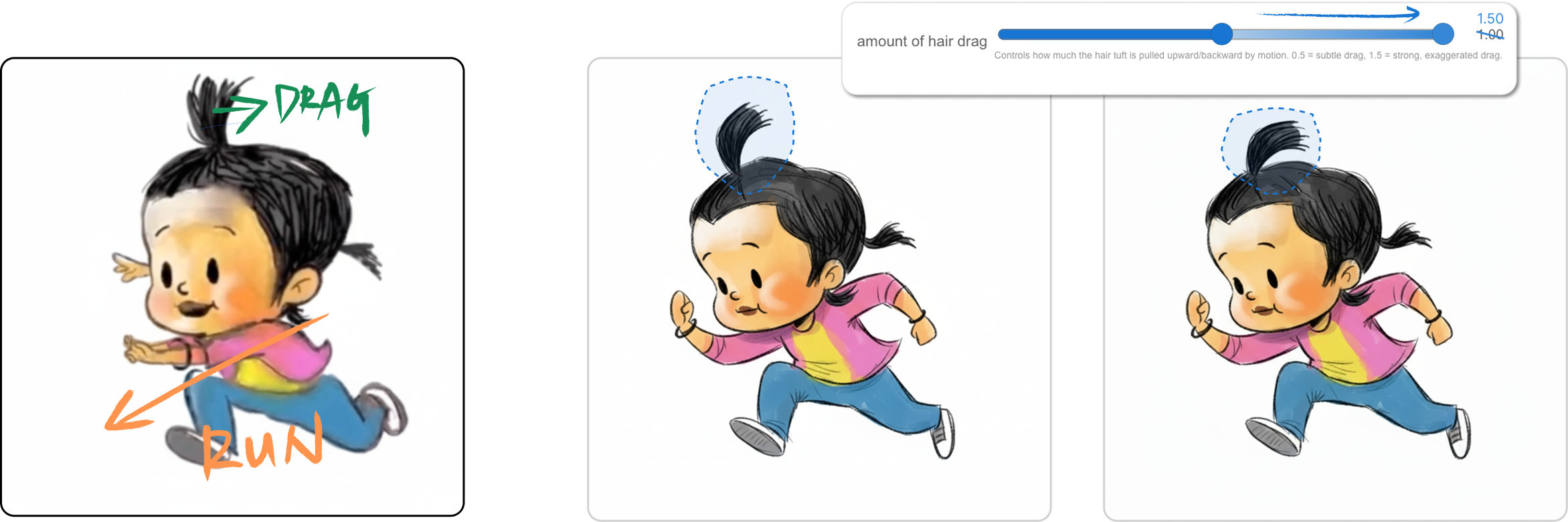}
    \caption{Demonstration of using dynamic slider to adjust the motion range for the ``DRAG'' motion.}
    \Description{Three panels show adjusting the ``DRAG'' motion range with a dynamic slider. Left: the character drawing is annotated with ``DRAG'' on the hair and ``RUN'' on the body. Middle and right: two generated running frames are compared while a slider labeled for hair drag is moved, changing how far the ponytail lags/trails behind the head.}
    \label{fig:usage_example_01_slider}
\end{figure}

This example illustrates a common consideration in animation design: a primary motion (running) accompanied by a secondary motion (hair drag), and demonstrates how to adjust the timing of their overlap.
Starting from the initial frame, as shown in Figure~\ref{fig:usage_example_01}-1, we place a long orange arrow labeled ``RUN'' to indicate the child's primary motion and direction of the movement. 
We then add a green arrow near the child's hair to denote the secondary motion: a slight delay that produces a follow-through drag. 
Next, we click Infer Motion (Figure~\ref{fig:usage_example_01}-2).
The system overlays two motion tags (Figure~\ref{fig:usage_example_01}-3) near the notations: ``body run'' and ``hair drag''; and hovering over a tag reveals more detailed information. 
The timeline is then constructed automatically, decomposing the two motions into seven tracks corresponding to individual primitive motions. 
The timing of the primary and secondary motion is often carefully overlapped, which we can specify by dragging and adjusting motion blocks on the timeline.
In doing so, we specified the next three keyframes to generate. 
Clicking ``Generate Frames'' button (Figure~\ref{fig:usage_example_01}-4) produces these three keyframes in sequence. 
Continuing to notate on the generated frames extends the sequence into a longer run of keyframes (Figure~\ref{fig:usage_example_01}-5).
Using the dynamic sliders, we can also adjust the range of the ``drag'' motion (Figure~\ref{fig:usage_example_01_slider}).


\begin{table*}[htp!]
\caption{The table records our interviewees' demographic information, including genders, range of ages, occupations, experiences in animation creation in years, and self-rated expertise in animation design in 5-item Likert scale.}
\vspace{-3mm}
\label{tab:interview_demographic}
\begin{tabular}{lcclcc}
\toprule
\textbf{ID} & \textbf{Gender} & \textbf{Age}   & \textbf{Occupation}  & \textbf{Experience} & \textbf{Expertise (1-5)} \\
\midrule
P1 & F & 26 & Creative Designer       & 4 years     & 4                  \\
P2 & M & 41 & Sr Technical Research Artist                & 20+ years      & 5                   \\
P3 & M  & 46 & Sr Quality Engineer & 10+ years        & 5               \\
P4 & M & 32 & Digital Artist       & 10+ years          & 5               \\
P5 & F & 26 & Indie Game Developer             & 8 years & 3                          \\
P6 & M & 45 & Designer       & 10+ years       & 5                   \\
P7 & F & 31 & Ph.D. student       & 2 years & 3                          \\
\bottomrule
\end{tabular}
\end{table*}

\paragraph{Object Animation}

\begin{figure}
    \centering
    \includegraphics[width=\linewidth]{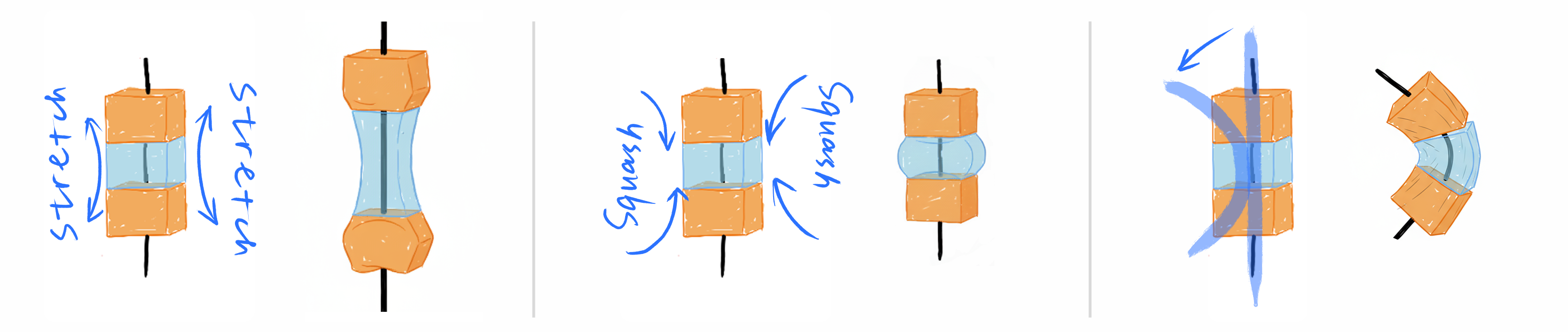}
    \caption{Object animation with \conceptname{} illustrating squash-and-stretch and shape deformation.}
    \Description{A row of small diagrams shows an object (a capsule-like shape) annotated and animated using notations. The sequence illustrates squash-and-stretch (the shape elongates and compresses) and deformation during motion, with blue handwritten cues and curved arrows indicating stretching, squashing, and bending.}
    \label{fig:usage_examples_cube}
\end{figure}

This example (Figure~\ref{fig:usage_examples_cube}) demonstrates how \conceptname{} conveys dynamism in static objects through shape deformation. 
We animate a stack of three cubes with coupled motion to show how notations can coordinate movement across multiple objects.
In the first two keyframes, arrows encode direction and squash-and-stretch. 
In the final keyframe, a broad overlaid stroke abstracts the transition from the initial to the target shape. 
In all cases, the VLM infers that cubes move together over time.

\paragraph{Particle System Dynamics}

\begin{figure}
    \includegraphics[width=\linewidth]{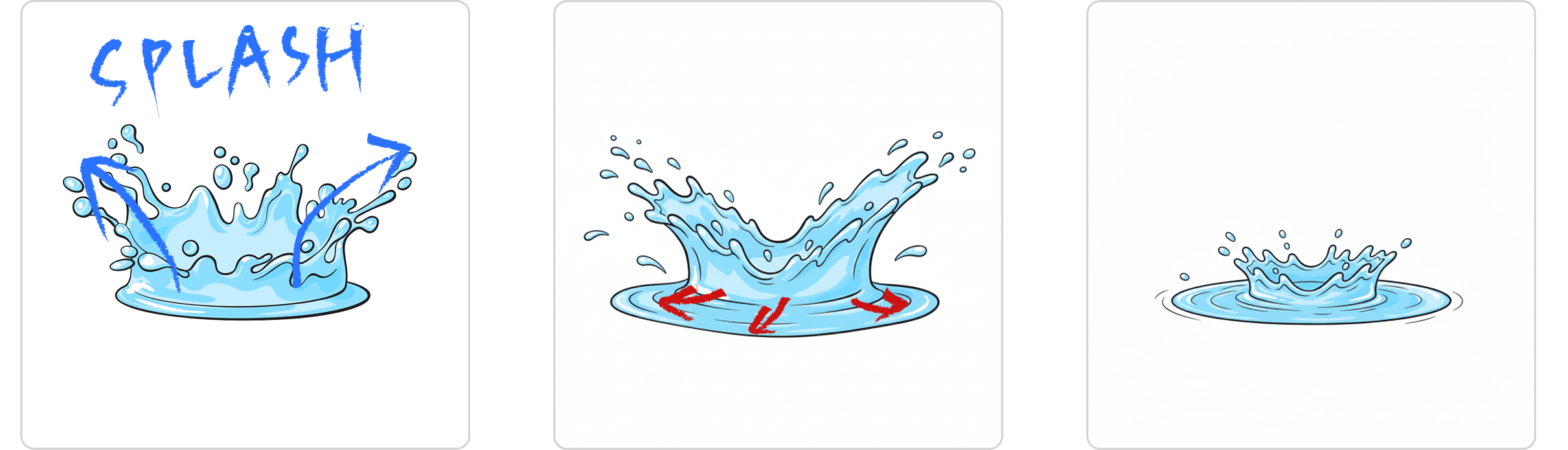}
    \caption{Demonstration of creating animation effects for particle system such as water with \conceptname{}.}
    \Description{Three panels depict a water-splash effect created with notations. Left: a splash drawing labeled ``SPLASH'' with arrows indicating outward motion. Middle: the splash expands and spreads with droplets moving away from the center. Right: the water settles into a smaller ripple/splash remnant.}
    \label{fig:usage_example_water}
\end{figure}

In this final example (Figure~\ref{fig:usage_example_water}), we use the notation to drive a particle system animation of a water splash. 
The three panels illustrate a temporal progression from the moment of impact (left), through the peak of the crown splash (center), to its dissipation (right). 
Directional arrows specify the intended motion field for particles: outward arrows around the rim indicate radial flow along the water surface, while upward arrows convey the vertical momentum that generates the crown. 
Arrow length and curvature encode both the expected trajectory and relative magnitude, which together suggest the emergent shape of the splash.
This example suggests the possibility beyond descriptive guidance, the notation can act as a lightweight control interface for physics-based effects. 
Interpreting arrows as velocity vectors allows a renderer to initialize emitter directions and speeds, while additional notations can parameterize features such as splash height, sheet thickness, and droplet spread.

\section{Expert Evaluation}
Using the developed prototype as a testbed, we conducted a user study with professional animators to evaluate the authoring experience of \conceptname{}. 
We aim to answer the following research questions: 1) how professionals use and perceive \conceptname{}; 2) how well a \conceptname{} system can support their control over the animation keyframe editing and how users can handle the misinterpretations; and 3) how might \conceptname{} impact user's way of thinking.

\subsection{Participants}

We recruited 7 experts in animation creation (3 females, 4 males). 
The detailed demographic information is listed in Table~\ref{tab:interview_demographic}. 
The studies were conducted remotely via Microsoft Teams. 
Participants accessed our system through a web browser using a pen-based input device: 2 were using iPad, 3 were using Movink, and 2 were using other tablets with the same functionality.
Each participant received a \$50 USD Amazon gift card for the 90-minute session.

\subsection{Task Design}
We designed two complementary tasks to examine how professionals use \conceptname{} and to identify limitations.

\begin{itemize}[leftmargin=*]
\item Constrained task (targeted effect):
Participants were given a starting static image and a brief natural-language description of a target animation effect, then asked to realize it using \conceptname{}. 
Prompts included: (\emph{i}) character: human celebrates a triumph; 
(\emph{ii}) animal: frog jumps to mid-air; 
(\emph{iii}) natural dynamics: water pours onto fire. 
We cover different animated subjects to examine how participants notate in different contexts.
\item Exploratory task (open-ended): Participants sketched their own initial frame and freely explored any animation they wished. 
This task aims to uncover unexpected behaviors and limitations.
\end{itemize}

\begin{figure*}[hbtp!]
  \centering
  \includegraphics[width=\linewidth]{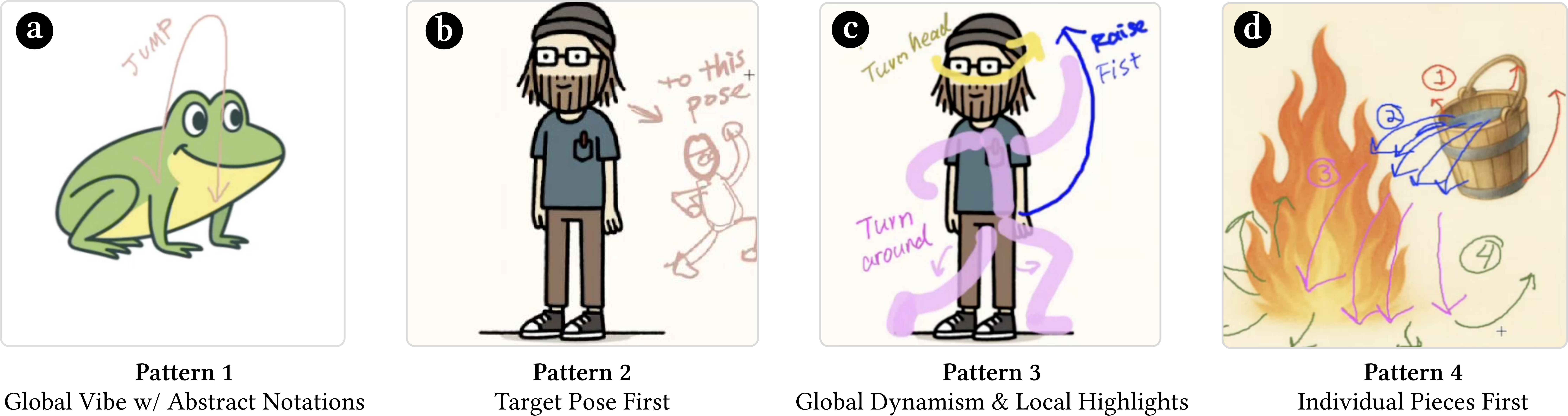}
  \vspace{-4mm}
  \caption{Representative examples of the four notating patterns identified in our content analysis. The patterns range from (1) expressing \emph{global vibes via abstract notations}; (2) defining the \emph{target pose} explicitly; (3) combining \emph{global dynamism with specific local adjustments}; to (4) breaking down the scene into \emph{individual pieces or sequential flows}.}
  \Description{Four example panels labeled (a–d) illustrate distinct notation patterns. (a) A frog labeled ``JUMP'' shows broad, abstract motion intent (global vibe). (b) A character sketch highlights setting a clear target pose first. (c) The same character is covered with large sweeping arrows plus small notes (e.g., turn head, raise fist), combining global dynamism with local highlights. (d) A campfire scene is annotated with multiple numbered arrows and paths, breaking motion into individual parts or sequential flows.}
  \label{fig:user_notation_examples}
\end{figure*}

\subsection{Procedure}
We began by briefing participants on the study purpose, procedure, approximate duration, and data collection practices, then collected demographics and background on animation design. 
Next, we introduced the concept of \conceptname{} and emphasized that participants were free to invent and use any notations. 
After a short walkthrough of \conceptname{}, participants completed a hands-on warm-up with a predetermined image unrelated to the study tasks. 
Once comfortable, they performed the two tasks in sequence while thinking aloud. Finally, participants completed a post-study questionnaire evaluating their experience with \conceptname{}, followed by a semi-structured interview. 
All sessions were audio- and video-recorded.
Participant-generated notational sketches, the transcribed think-aloud data and interview data were analyzed through inductive thematic analysis.

\section{Findings}
We report the findings in several themes: overall impressions of \conceptname{}, notating patterns, failure cases encountered, how animators handle misinterpretation, how they make exaggerations and how their mental model shifts.

\subsection{Overall Impression}
Overall, our participants found this interaction concept is \emph{impressive}, \emph{useful}, and \emph{intuitive}. 
On a 7-point Likert scale evaluating the statement \textit{``It was easy to express my animation intent.''}, responses included: 2 strongly agree, 4 agree, and 1 neutral. 
Some of user-produced keyframes are shown in Figure~\ref{fig:user_study_results}.
P6 reflected, \qt{I think this is really exciting and fun ... It was fluid, you know, animation is such a time consuming process and things like this get me excited about how much time they could save.}
Participants appreciated the intuitiveness of \conceptname{} because they did not face a steep learning curve and felt their intention can be mostly captured, \pqt{it was more like open to interpretation and like a quick and easy thing to do and that felt like I didn't have to learn any new language to do it. Such a way [notional animating] is approachable and exciting to me. I just draw arrows and things and write things on the screen and it seemed to interpret most.}{P3}

Participants emphasized that \conceptname{} felt intuitive because it aligns with how they already think and communicate through sketching. 
P4 noted that, \qt{annotations are already part of how we work, when we revisit a piece or give someone direction, we mark things up to make the idea clear. It's how we visualize what to change.}
In contrast, they described prompt writing as a separate, learned skill that they had to learn \pqt{its own language which is very different from ours.}{P4}
This perspective aligns with how artists are trained to develop motion progressively through sketching: \pqt{The way we're taught is to start simple, block it out with basic lines and shapes, build the structure, then layer in detail, and in this tool, I'm doing things in this flow.}{P7}
Taken together, these comments suggest \conceptname{} supports a natural interaction approach that builds on animators' existing workflow.

\subsection{Notating Patterns}

\subsubsection{Observed notating patterns across participants}
\begin{figure*}[htp!]
  \centering
  \includegraphics[width=\linewidth]{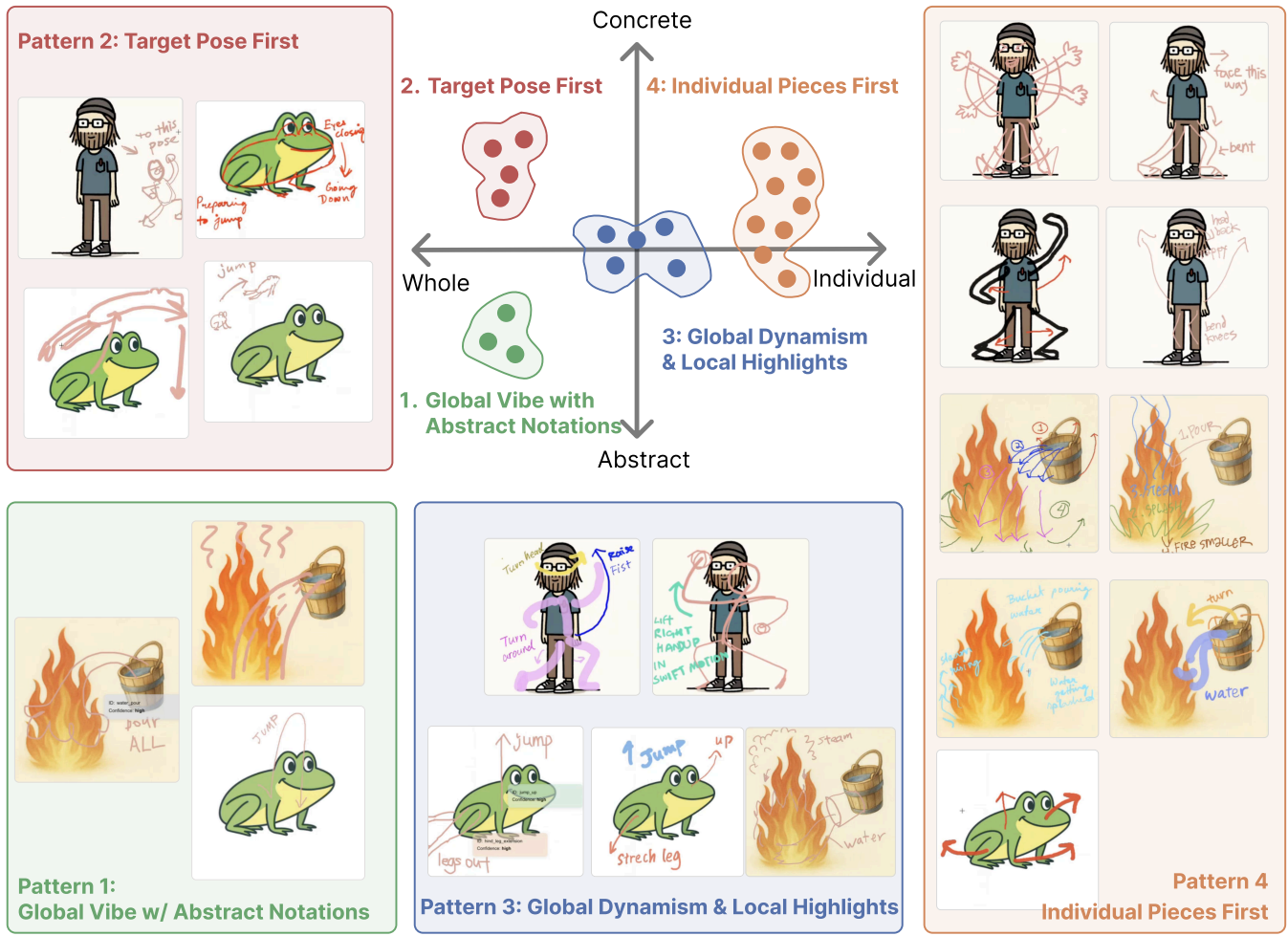}
  \caption{Identified four notation patterns spanning on the axis of the approach (abstract \vs concrete) and the focus (whole \vs individual). Pattern 1 is ``Global vibe with abstract notations'', pattern 2 is ``Target-pose first'', pattern 3 is `` Global dynamism first, then local highlights'', and pattern 4 is ``Individual pieces first''. }
  \Description{A quadrant diagram maps four notation patterns along two axes: abstract - concrete (vertical) and whole - individual focus (horizontal). Each quadrant is labeled with a pattern: (1) Global vibe with abstract notations (bottom-left), (2) Target pose first (top-left), (3) Global dynamism then local highlights (center/bottom), and (4) Individual pieces first (top-right) with example annotated drawings (frog, character, and campfire/bucket scenes) shown in colored boxes around the plot.}
  \label{fig:user_study_notating_patterns}
\end{figure*}

We observed four recurring notating patterns from participants. 
These patterns span along two axes: \emph{focus} (whole body or scene \vs~ individual parts) and \emph{abstractness} (abstract intention \vs~ concrete specification). 
We regard notations (\eg, arrows) as more abstract while the user-drawn pose as more concrete. 
The same animator may switch patterns across tasks or even try different ways within a single task, we list how participants adopt different patterns in Table~\ref{tab:patterns_adopted}.
Despite their different notating patterns, all animation intentions can ultimately be captured within our formalized representation.
Below we describe each pattern with one representative example, the landscape of notation patterns can be found in Figure~\ref{fig:user_study_notating_patterns}.

\begin{table}
  \centering
  \vspace{2mm}
  \caption{Notating patterns adopted by each participant.}
  \label{tab:patterns_adopted}
  \begin{tabular}{l p{.5\linewidth}}
    \toprule
    \textbf{Participant} & \textbf{Patterns Adopted} \\
    \midrule
    P1 & Pattern 1, 3, 4 \\
    P2 & Pattern 1, 2 \\
    P3 & Pattern 3, 4 \\
    P4 & Pattern 2, 3, 4 \\
    P5 & Pattern 1, 4 \\
    P6 & Pattern 3, 4 \\
    P7 & Pattern 4 \\
    \bottomrule
  \end{tabular}
\end{table}

\textbf{Pattern 1: Global vibe with abstract notations.}
This pattern reflects a process-oriented mental model where animators begin at a high level, focusing on \emph{how the motion unfolds}.
They typically sketch the overall motion arc or energy (via \textsc{Path}) and add short semantic labels (\eg, \textit{jump}).
\textsc{Target} is often omitted and \textsc{Source} is implicit (\eg, \textit{the whole frog}). 
One notation often express multiple intended keyframes. 
For example, as shown in Figure~\ref{fig:user_notation_examples}.a, P6 use the arrow which consists of a small downward arc and a huge upward and downward arc to indicate \qt{... (the frog) would be getting down ready for a very high leap, and then hitting the ground again then kind of back up into this main position.}

\textbf{Pattern 2: Target pose first.}
This pattern reflects a state-oriented mental model where users focus on \emph{what the next key pose should look like}. 
They explicitly draw the desired \textsc{Target} pose for the whole character (\eg, Figure~\ref{fig:user_notation_examples}.b), while leaving the transition dynamics (\emph{how} to reach that state from the current one) implicit. 
This approach reflects a preference for direct artistic execution, as emphasized by P2: \qt{Let animators draw! ... the overall look is in my mind and naturally comes to my hand.}

\textbf{Pattern 3: Global dynamism first, then local highlights.}
This pattern reflects a hierarchical mental model where users establish the overall motion dynamism before refining local details. 
They typically begin by notating the overall action using thick strokes, then switch to thinner strokes to annotate specific parts for detailed adjustments (\eg, \textit{bend the knee}, \textit{turn head}), as shown in Figure~\ref{fig:user_notation_examples}.c.
In other words, a global \textsc{Target} for the whole plus selective, part‑level constraints to specify the \textsc{Path}.
As P3 explained, \qt{that's actually a big part of animation is you get the central line first and that informs all the rest of your limbs.}

\textbf{Pattern 4: Individual pieces first.}
This pattern reflects a compositional mental model where users decompose complex motion into primitive units and annotate each part separately.
Participants leveraged mixed strategies within this pattern, sometimes using abstract notations to specify the \textsc{Path} for specific limbs, or defining a concrete local \textsc{Target} pose for others.
They also use color coding to distinguish parts and numerical labels to coordinate timing, \eg, Figure~\ref{fig:user_notation_examples}.d.
However, this granular approach often resulted in visual clutter within the fixed canvas.
As P4 noted, \qt{(when notating,) I felt like restricted in that (fixed) size, being able to zoom in and zoom out might help.}
Although the observed examples typically showed notations for each part appeared together, feedback suggests this was likely an artifact of system latency.
P3 explicitly described a desire for an iterative, layered workflow: \qt{if this worked faster... I would be doing this in several steps... turning his head first and then turning his body... a whole bunch of staggered individual motions.}

\subsubsection{Design implications from notating patterns.}
\label{sec:implications_from_notation_patterns}
Collectively, these four patterns validate the core premise of \conceptname{}: rigid vocabularies cannot accommodate the varied mental models of animators, confirming the necessity of \emph{user-defined abstractions}. 
Beyond this validation, the observed behaviors offer critical design insights regarding the \emph{agency} of AI models and the \emph{structure} of animation representation. 
Regarding the AI agency, the contrast between the abstract \emph{Global Vibe} (Pattern 1) and the  \emph{Target-Pose} (Pattern 2) reveals a dynamic spectrum of control: users shift between offloading low-level planning to the model (acting as a \emph{co-creator} in Pattern 1) and demanding strict adherence to geometric constraints (acting as an \emph{executor} in Pattern 2). 
Thus, future systems could infer the \emph{intended level of AI agency} from the notation's abstractness. 

Furthermore, these notating patterns also point to potential improvements in the \emph{structure} of animation representations.
While Pattern 4 (\textit{Individual pieces first}) reflects a compositional structure that aligns with our current representation where all animation units are treated equally, Pattern 3 (\textit{Global dynamism with local highlights}) suggests the need for a hierarchical structure. 
Future representations could explicitly distinguish between \emph{global} constraints and \emph{local} adjustments so that models can prioritize global dynamism during generation while ensuring that local details complement the primary motion rather than conflict with it.

\subsection{Shortcomings in Notations and Models}
The observed failures could be attributed to three factors: limitations in notation expressiveness, the robustness of VLM's reasoning, and the instruction-following ability of image generation models.

\subsubsection{Expressive affordance of notations.}
Aligning with our content analysis (Section~\ref{sec:design_space}), all participants commented that it was easy to communicate \emph{2D spatial} \emph{and geometric} information through notation: how the mover should be oriented, where it should move, how it should look (\eg, the pose), how its volume (size) should change (to reflect stretch), and what it should feel like in animation (\eg, squashy and stretchy).
However, a fundamental limitation emerged regarding \textit{3D spatial changes} (\eg, twist, tilt, rotate) and \textit{fine-grained occlusion}.
Participants (P1, P2, P3, P5, P7) expressed their struggle to flatten complex 3D rotations onto a 2D canvas. 
It should be noted that this friction lies in specifying \emph{how it rotate in a specific way}, not the \emph{semantic intent of rotation}.  This explains why writing text does not help in such cases; articulating a precise 3D trajectory capturing the specific arc, tilt, and velocity of a complex turn is as geometrically ambiguous in natural language as it is in flattened 2D notations.
P3 gave an example where it is hard to specify \qt{... his head moves up while turning around, and then it moves like up and down.} 
Additionally, \textit{fine-grained local motions with occlusion} (\eg, finger curls) are hard to notate legibly without clutter. 
P2 noted that \qt{the marks either became too crowded to read or hard to capture the subtle motions and their dependencies}. 
P2 suggested that such spatial nuances are best captured not by writing or drawing, but by performance: \qt{the best way to express this is the gestural input... how I myself perform it.}
Finally, \textit{non-geometric transforms (\eg, color, lighting)} are important for animation rhythm and mood but are not currently considered in \conceptname{}. 
P1 also suggested it is possible to indicate \emph{lighting} and \emph{color} with notations.

\subsubsection{Interpretation precision and generative fidelity of AI models.}
\label{sec:model_shortcomings}

\begin{figure*}[htp!]
  \centering
  \includegraphics[width=.95\linewidth]{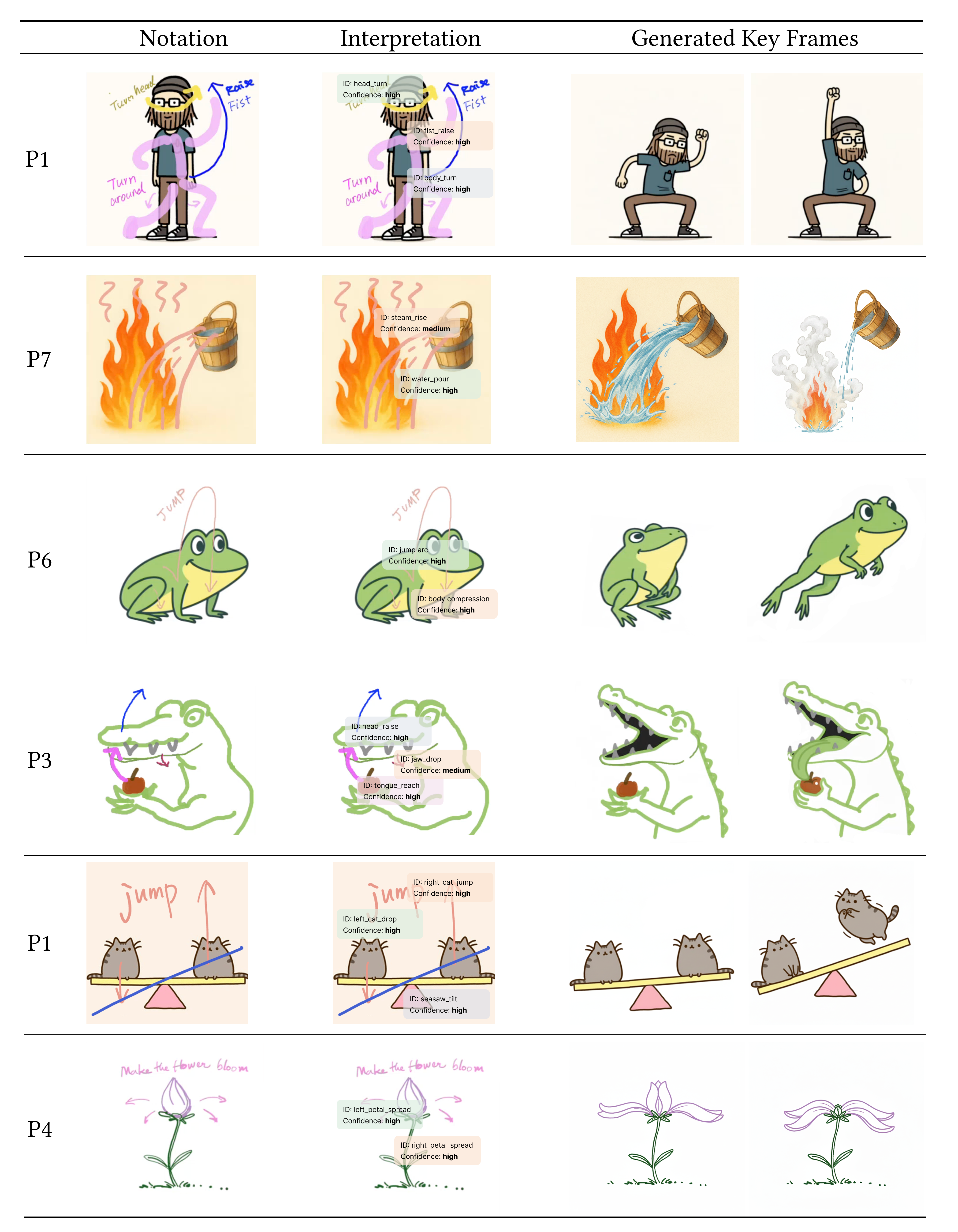}
  \caption{Participant outputs from the expert evaluation: the first three results are from the structured task; the last three were produced during the open-ended exploratory task.}
  \Description{A multi-row results table with three columns: Notation, Interpretation, and Generated Key Frames. It shows participant-generated examples. In each row, a hand-drawn scene with colored motion marks (notation) is followed by the system's parsed motion tags with confidence, and then two generated keyframes illustrating the motion (e.g., a character turning and raising a fist, a bucket pouring water to affect a fire, a frog jumping, an alligator moving head/tongue, cats jumping on a seesaw, and a flower blooming as petals spread).}
  \label{fig:user_study_results}
\end{figure*}

In most cases, the vision-language model can interpret users' intention to a reasonable degree; on a 7-point Likert scale evaluating the statement \textit{``The AI correctly understood my intent.''}, responses included: 2 agree, 3 somewhat agree, and 2 neutral. 
However, even when the VLM correctly interprets user intent, the downstream image generation model sometimes fails to execute the instruction faithfully, exhibiting three primary failures: \emph{(i) spatial mismatch, (ii) deformation resistance, and (iii) stylistic drift}. 
First, the model sometimes mess up the character's left/right or viewer's left/right, resulting in orientation errors like the unintended flipping in P1's character animation (Figure~\ref{fig:user_study_results}). 
Second, the current image generator often struggles to apply exaggerated shape deformation (\eg, stretch, compress, overshoot) to subjects it perceives as rigid; for instance, in Figure~\ref{fig:user_study_results}, P6's frog remained a fixed volume despite instructions to compress, whereas P1's cat successfully deformed to some degree. 
This inconsistency suggests a need to fine-tune models to prioritize explicit deformation instructions over implicit physical realism. 
Finally, we observed style drift in some cases where the model altered the artistic identity of the base drawing, such as P3's crocodile losing its specific ``goofy'' aesthetic but being altered to a more generic style.

\subsection{Handling Misinterpretation}

\subsubsection{Understanding AI's interpretation.}
When evaluating the statement of \textit{``I received sufficient feedback to know where and what to correct when the AI misinterpreted my intent.''}, the response included 3 strongly agree, 2 agree, 1 somewhat agree, and 1 neutral.
We observed two feedback–checking styles. 
Most participants adopted a cautious, incremental approach by verifying the system's feedback at each step before committing to full generation; whereas one participant (P2) preferred to ignore intermediate feedback, generate first, and only then inspect the feedback.
Overall, participants found the in-place motion tags helpful for quickly revealing how the model parsed their notations. 
P5 commented on the $\langle source, path, target \rangle$ structure: \qt{... the first three is enough to give me a quick overview of like what it's saying.}
Participants also asked for richer transparency about how secondary notations were interpreted. 
As P3 suggested, \qt{If it [the system] could also tell me how it interpreted my different usage of brush sizes and colors, that would be a big help, because if something goes wrong and I know better its thinking process then I know how should I change my notations.}

\subsubsection{Correcting the misinterpretations.}
When the system misinterpret user intent, participants most often corrected it by iterating on their notations. 
More than half of the participants (4/7) opted to add \emph{secondary notations} as their immediate reaction to misinterpretations, utilizing strategies such as brief text labels to define motion semantics or color-coding to distinguish animation units. 
The remaining participants (3/7) chose to refine their \emph{primary notations}, for instance, by redrawing arrows or clarifying motion paths.
Occasionally, they corrected the interpretation directly by editing motion tags, but this was less preferred, only when iterating on notations keep failing. 
As P7 commented, \qt{I don't like switching between drawing and typing}.
This explains why most participants chose to iterate on notations rather than editing directly on the feedback. 
Finally, when elements were omitted, participants typically fixed this by adding new motions on the timeline blocks to enforce the intended actions.

\subsection{Making Exaggerations}
Exaggeration distinguishes animation from film; it makes animation believable and often humorous. 
For an interaction approach for animation keyframe authoring to be truly useful, it must allow animators to easily express and realize exaggeration. 
In our study, we explored how much \conceptname{} currently affords exaggeration across two axes: \emph{motion} and \emph{timing}.

\subsubsection{Motion (pose, deformation, range).} 
Exaggeration often lives in dynamic posing and shape deformation; being able to specify and adjust the range, intensity, and shape of motion is crucial. 
We observed that participants used \emph{secondary notations} such as stroke thickness to signal stronger deformation; for the frog leap, P4 said, \qt{So stretch will be like that. But I need to be very much. So I need to exaggerate that arrow also, so I'm changing to a larger brush size.}
Participants also frequently utilized dynamic sliders to refine motion range and intensity after the initial generation.
On average, each participant engaged with these widgets approximately 8.1 times throughout the session (\md{8.0}, \iqr{3.0}).
However, usage frequency varied notably by animation task.
Interaction was most frequent during the \emph{open-ended} task (\md{3.0}, \iqr{1.5}) and the \emph{frog leap} task (\md{2.0}, \iqr{1.0}).
In contrast, the \emph{fire and water} task saw minimal GUI intervention; two participants (P1 and P6) relied entirely on their notations, while the remaining participants used the sliders only 1.2 times on average (\md{1.0}, \iqr{0.5}).
P3 commented, \qt{I'm happy that I can have that slider for sort of control, animation is all about control to make it real. You know the rate of how this fire diminishes and how far that bucket pours and rotate.}
On the other hand, some participants noted that while dynamic sliders offer quick, high-level adjustments toward an exaggerated feel, they are still less direct than traditional direct manipulation in After Effects; as P6 put it, \qt{it's still less direct and less control over things. It's like suggestion versus direct manipulation.}

\subsubsection{Timing (event order, primary \vs secondary motion).}
Exaggeration also depends on coordinated timing such as anticipation, follow-through, and overlapping action. 
As P4 put it, \qt{things (in animation) actually happen in a natural way. The motion happens with some follow through or some things happen first and resulting to that secondary motion happens. So like if this guy is doing this, he will first bend down and then push, that sort of timing made the exaggerated flavor of animation.}
We observed two strategies: some participants added numeric labels (\eg, ``1'', ``2'') near marks to indicate event order; others kept the order implicit and expected AI to infer from the context, P2 noted \qt{it (the system) should know from the physics}. P6 was impressed when the model inferred anticipation before the leap: 
\qt{it was really impressive what it was able to infer of, this is what I was trying to do with the frog balance. Where it does do the compression first before stretching out for the leap as the anticipation. It's thinking about multiple steps, I think it comes down to a matter of control that is exactly animation needed.}
When AI's timing inference fell short (\eg, all decomposed motions landing together), participants highlighted the value of incorporating the timeline in \conceptname{} so that they can directly manipulate it to stagger and overlap actions. 
As P3 commented, \qt{it's really impressive I was able to use notations first then down to the timeline because that's really where you get interesting. Animation is like the the overlapping of various actions.}
They also valued per-object tracks, P5 recounted, \qt{because when you perform an action, multiple stakeholders will be affected and in this case it will be the fire, the water and the bucket itself. So it makes perfect sense to have three different tracks for these separate objects to define.}
Moreover, participants often iterate a lot across notations and the timeline: \pqt{that right feel is not always directly happens, but it happens gradually. You keep on iterating and changing a bit notations and drawing, adjusting the timing, and until the motion and that volume comes up correct.}{P4}

\subsection{Mental Model Shifts}
We wondered how \conceptname{} influences animators' mental models compared to their daily tools, and found two major shifts in how they think and work.
Accordingly, many participants suggested integrating \conceptname{} into their current workflow.

\subsubsection{From low-level sophisticated control to high-level animation vibes.}
When asked to compare their prior experience in familiar tools with \conceptname{}, all participants described a shift from tweaking parameters toward steering the \emph{vibe} or \emph{feel} of motion at a higher level. 
P6 described their thinking in \conceptname{}, beginning with: \qt{It's got squash, it's got weight, it's got gravity. It's talking about these things that I care about.} 
P3 contrasted authoring in After Effects with \conceptname{} and anticipated a near-term shift toward \conceptname{}-style interaction model:
\pqt{I know we're just about that time for change, so hopefully this will go fast. 
A lot of features that I've been using in After Effects lately were designed specifically to do this sort of tweaking around because in After Effects it's very easy to author keyframes directly. 
But this way here [notational animating] was introducing a lot of possibilities to kind of like distribute those key frames over an amount of time or sometimes distribute the values over an amount of time so that people can kind of just grab on to the totality and the vibe of the animation instead of stacking all stuff manually.}
P5, an independent game designer who often codes animations, reflected: \qt{when I program it in a loop, I'm just moving X–Y coordinates. Annotation feels more direct for me because I'm literally just put down the motion and the feel and all what I want with the pen. It helps me think and feel it intuitively, compared to that variables.}
P7 commented that, \qt{sort of like blending the brainstorming sketching phase in the animation phase. It's just like you're kind of vibing it out.}

\subsubsection{From doing ``one thing at a time'' to ``multiple things together''.}
Animators in classical workflow typically do ``one thing at a time'': \pqt{we normally do an animation where all these things are different pieces and animate one at a time, test it until it works, then move on to the next, then stagger them together.}{P6}
However, in our study, we observed participants often choose to animate multiple things together.
In the water and fire example, we observed most animators \emph{notate} bucket, water, fire, and steam altogether and leave AI to \emph{infer} the timing and their dependencies.
P1 explained that \qt{there's some interconnection between these things. These are all kind of they're coming together as a system. So that's pretty cool that I was able to work on them altogether.}
Similarly, in the frog example, P6 used one notation to indicate the whole process that the frog first compress to prepare the jump then jump a big leap arc. They reflected that \qt{I think of it all as one thing and all together. I love that I was able to tell a specific order of events, like use that little down thing and the big arc to specify the anticipation at the beginning before the jump, and I'm impressed that it was able to figure that out. I'm really interested to use it to see how multi-frame animation comes together.}

\subsubsection{Integrating \conceptname{} into current workflow.}
Participants showed great interests in getting notational input plug into established pipelines to handle both high-level ``vibe-first'' guidance and low-level precise edits. 
On the one hand, P6 envisioned a hybrid model inside After Effects: \qt{bringing this kind of system into AE feels like the best of both worlds because it still gives me the full control and precision I'm already used to. Wouldn't it be cool if there were this additional things that could let me quickly try things out and potentially fill in some of the gaps of things I don't know how to do as well.}
On the other hand, besides rough and quick experiments, P2 also suggested \qt{notations can also be used for very minor adjustments}. 
As P4 noted, \qt{it can be useful when they just want to tweak a final animation they have created, like okay, this keyframe here I just want this to move a bit here}.
Participants also anticipated different adoption curves. 
P4 cautioned that, for professionals, \qt{that might take some time to totally get into their pipeline. All the professional who are already in the industry, they are very accustomed to their own workflows, the changing will be gradual; but this will be quite interesting with the new generation of GenAI artists who are trying out new ways of telling story and motion.}

\section{Discussion}
We regard this paper as an initial step towards realizing \conceptname{}.
Overall, our qualitative findings show that the core interaction is intuitive and useful for animators: it draws on their familiar, conventional sketching and requires no learning.
With our prototype, participants were able to express a range of animation nuances, including multiple forms of exaggeration, and were often impressed that most informal notations can be correctly interpreted into their intended animation.
Collectively, these results support \conceptname{} as a promising paradigm for higher-level animation keyframe authoring which has the potential to serve as a complement (but not a replacement) to classic GUIs (\eg, After Effects), which excel at fine-grained, low-level precision control.

\subsection{Reflections and Future Directions}

Despite participants feeling they can express most of their intent with \conceptname{}, there is still much room for improvements. 
We see the opportunities and needs for expanding the notations.
For example, camera movement is conventionally expressed through marks, and prior work has examined arrow-based notations for camera motion in storyboarding. 
Participants also suggested extending notation to non-geometric transformations (\eg, color and lighting). 
Accommodating these additions will require extending our current formal representation of animation intent. 
We designed this representation to be extensible so new dimensions can be accommodated without breaking the existing structure.

We also identified some bottlenecks related to current models to make \conceptname{} more useful. 
First, the intention inference and keyframe generation should be faster.
A tight feedback loop encourages bold exploration; high latency will push users toward conservative choices. 
Moreover, what matters for professional adoption ultimately is the fidelity of the generated keyframes. 
Although optimizing output quality was not our focus here, it is decisive for whether \conceptname{} can shoulder production work or remains primarily an ideation tool for quick, rough exploration. 
We are optimistic that near-term model advances will be able to make progress on interpretation robustness and generative fidelity (as discussed in Section~\ref{sec:model_shortcomings}) to enable \conceptname{} to mature into a serious production approach.
Additionally, our current framework lacks a dedicated mechanism to prevent the VLM from imposing unsolicited creative intent when users require strict control. 
As discussed in Section~\ref{sec:implications_from_notation_patterns}, the abstractness of a user's notation serves as a strong signal of their intended agency. 
Future systems should leverage this signal to dynamically regulate the model's behavior, specifically inhibiting the tendency to ``polish'' or restyle the animation when users demonstrate a preference for explicit, low-level control.
We hope this work can help bridge the community of HCI and AI, calling out the needs for understanding artists' vision such as principles (\eg, anticipation, follow-through, exaggeration) which go beyond real-world physics, so models can reason about and generate the intented animation faithfully.

Beyond keyframes, the ``last mile'' we envision is automatically rendering in-between frames from the produced keyframes (most likely via video models).
Although it is feasible to do so with the state-of-the-ard models, results are not yet production-ready, and current video generation remains slow. 
Until generation is fast enough for interactive playback during design, animators cannot easily use full animated previews to test and refine their choices.
Nevertheless, we remain optimistic that rapid advances in video models will make this practical in the near future.

\subsection{Animation Representations across Different Abstraction Levels}
Central to our \conceptname{} is the formalized animation intention representation.
If we look back to the history of the animation creation, each interaction model corresponds to a representation for animation. 
Back to the classical hand-drawing era, animation representation was abstract, artist-centric constructs in simplest visual form (\eg, line of action) where artists start from the simplest forms and get the dynamism of animation and iteratively add details on top of it to get the final animation. 
Later in the digital animation era, the representation shifted to concrete, lower-level primitives: per-keyframe translation, rotation, scaling, and physical simulations. 
Interestingly, this sharp change of the representation from the abstract to the concrete one is not really aligned with how artists think about the animation, even today, master animators still tend to think abstractly. 
This separation likely explains why animation production stages are distinct, from initial freeform sketches to final production software ~\cite{jahanlou2021challenges}.

Recently, with AI's advancements, it is promising to offload more low-level tedious labor to AI, and our focus in this paper is to facilitate a direct manipulation interaction that accommodates different levels of abstraction of their intent.
We want to highlight the way in \conceptname{} that adding a middle-layer representation (our formalized animation intention representation) bridges the fluid, sketch-like notations that artists prefer and the implicit space AI works on. 
As a formal layer between human-facing notation and AI embeddings, it aims to be extensible and flexible, aligning abstract intent with concrete execution. 

Our goal is not to define a single best representation for AI-driven animation authoring. 
Instead, we aim to open a design space and invite the community to explore feasible higher-level representations that align with how animators think and work. 
We approach the problem from a \emph{movement-centric} perspective, \ie, ``what aspects about movements users wish to specify''; but other ways such as \emph{effect-centric} representation are also promising, \eg, reifying animation principles or physics-driven effects as first-class instruments~\cite{Kazi:2016:Motion-Amplifiers,Xing:2016:Energy-Brush}.
Style is also central to animation (\eg, manga, Pixar, Disney). 
Future work could let users define and reuse personal style conventions as part of their notations. 
More broadly, it remains open how such notations evolve over time with cultural practices.
We hope \conceptname{} is a step toward representations that are both human-legible and AI-operable, blurring the boundary between ideation and execution in animation authoring.

\subsection{Limitations}
\label{sec:limitation_and_future_work}
Our study recruited only professional animators. 
This study design prioritized understanding how \conceptname{} fits into professional practice and its \emph{expressive ceiling} in a domain governed by nuanced principles (\eg, anticipation, follow-through, exaggeration). 
As such, experts are better positioned to set demanding goals, articulate complex intentions, and judge whether this interaction approach helps make motion ``a certain way'' rather than merely making things move. 
Their proficiency with GUI-based tools (\eg, After Effects) also enabled sharper contrasts between interaction paradigms.
We anticipate that \conceptname{} will also benefit novices, though their perception and usage patterns may differ from those of professionals. 
Professional animators possess internalized knowledge of implicit notation conventions gained through training and practice. 
In contrast, novices lacking this exposure may face a \emph{cold start} problem, unsure of how to initiate the interaction. 
A possible solution is to offer a set of \emph{default} notations or generating context-aware ambient suggestions. 
These aids would help novices discover \emph{what} can be notated and \emph{how}, eventually preparing them to invent their own notations. 
Evaluating the learning curves of novices remains a critical direction for future work.

Our current evaluation is qualitative in nature, focusing on understanding how professionals utilize \conceptname{} within our prototype.
Future quantitative measurements could deepen our understanding of \conceptname{}. 
Although \conceptname{} is grounded in real-world sketching practice, it remains unclear how closely users' digital notations match what they would sketch on paper. 
We hypothesize that notation behaviors shift with both the medium (static paper \vs~ an interactive system with feedback) and the audience (self-reflection \vs~ communicating with an AI model). 
To test this, a comparative study could ask participants to express the same animation intentions using both paper and \conceptname{}, enabling us to quantify systematic differences in notation.
Finally, real animation projects often span days to months. 
Although our in-lab study included an open-ended exploration session that let participants experiment with their own drawings and any effects they wished, the tasks remained toy-scale relative to production work. 
A longitudinal deployment study on real projects will therefore be invaluable for assessing the broader, real-world impact of \conceptname{}. 
\section{Conclusion}
In this paper, we introduce \conceptname{}, an interaction concept for animation keyframe authoring that reifies sketched notations as animators' intentions. 
These notations guide the creation and editing of keyframes, which can subsequently be consumed by automatic methods (\eg, recent frame-interpolation models) to generate animation frames. 
Through content analysis and expert interviews, we characterize the notations' informal nature: non-categorical, ambiguous, and combinatorial. 
We then formalize them into a structured representation of animation intent, enabling recent vision-language models to interpret the notations systematically. 
As an initial step toward realizing \conceptname{}, we built a prototype featuring a two-level feedback mechanism and dynamic UI widgets that provide fine-grained control complementing the high-level notations. 
In a preliminary expert evaluation, we examined how animators perceive and use \conceptname{}. 
Our findings suggest that sketched notations can effectively support animation authoring and inform future research on AI-assisted animation tools that preserve artists' agency.

\begin{acks}
We sincerely thank Seth Walker, Val Head, and Wilmot Li for their insightful discussions during the early stages of this project, and the anonymous reviewers for their invaluable suggestions, which greatly improved this work.
We also thank Crystina Zhang for her kind help with several figures. 
Finally, we are grateful to all our participants for their time and valuable input.

This work is supported in part by the Natural Sciences and Engineering Research Council of Canada (NSERC) Discovery Grant (\#RGPIN-2020-03966), the Canada Foundation for Innovation (CFI) John R. Evans Leaders Fund (JELF) (\#42371), and a gift fund from Adobe.
We acknowledge that much of our work takes place on the traditional territory of the Neutral, Anishinaabeg, and Haudenosaunee peoples. 
Our main campus is located on the Haldimand Tract, the land granted to the Six Nations that includes six miles on each side of the Grand River.
\end{acks}

\bibliographystyle{ACM-Reference-Format}
\bibliography{references.bib}

\appendix
\section*{Appendix}
\label{sec:appendix}
\section{Prompt for Interpreting Notations}


\begin{aptdispbox}
\acmcallout{Prompt for Interpreting Notations:}{\textbf{Role \& Goal:}\par
Interpret raster sketches with motion marks. For each mark, identify its role, group marks or decompose a mark (only when necessary) into motion units, and return a structured description using <source, path, target> plus relevant secondary modifiers.

 \textbf{Conceptual Model:}
\begin{itemize}[leftmargin=*]
  \item Motion unit: one intended motion.
  \item Primary tuple <Source, Path, Target>
  \begin{itemize}
    \item Source: the moving part(s).
    \item Path: movement type and direction.
    \item Target: intended end state.
  \end{itemize}
  \item Secondary modifiers: visual properties, text/letters/numbers, emphasis, symbols.
  \item Prioritize role over form. Allow incomplete units; infer missing pieces when intent is clear.
\end{itemize}

\textbf{Grouping \& Association:}
\begin{itemize}[leftmargin=*]
  \item Cluster by proximity, common style (color/thickness/texture), and shared role/direction.
  \item Associate nearby numbers/letters/text with the most relevant path or target.
\end{itemize}

\textbf{Grid:}
\begin{itemize}[leftmargin=*]
  \item 30x30 cells; origin at bottom-left (0,0); x-> right, y-> up.
  \item Use integer intersections 0-30 and .5 for edges/centers.
\end{itemize}

\textbf{What to Extract (per motion unit)}
\begin{itemize}[leftmargin=*]
  \item color: unique per unit.
  \item roi\_bbox: [x\_min, y\_min, x\_max, y\_max] in grid units.
  \item primary: \{ source, path, target \} — concise, consistent with the model above.
  \item secondary\_modifiers: entries of \{ property: color|thickness|text|number|letter|style|other; value: "..."; intended\_meaning: "..."; scope: source|path|target|unit \}.
  \item temporal\_order: integer if deducible from nearby numbers/letters; else null.
  \item confidence: 0-1.
  \item natural\_language\_summary: 1-3 sentences describing the intended move and end state (where it lands, contacts/near-contacts, rough orientation/pose); include 2-4 feel adjectives (e.g., fluid, snappy, heavy, brisk).
  \item unassigned\_marks: brief notes with small grid bboxes for marks not confidently assigned.
  \item global\_timeline: ordered list of unit ids if numbers/letters imply a sequence.
  \item legend\_inferred: propose consistent color/style meanings if patterns emerge (e.g., "red = paths").
\end{itemize}

\textbf{Dimensionality Controls (required)}\par
Provide 1-3 neutral sliders, default 1.0, grounded in marks:
\begin{itemize}[leftmargin=*]
  \item Amplitude/geometry (pick 0-2): e.g., height of raise, reach length, arc sweep amount.
  \item Directional bias (<= 1): e.g., forward bias of the swing. (0.5-1.5)
  \item Timing/Energy (when cued by echoes/weight): e.g., tempo/quickness.
\end{itemize}

\textbf{Output Format:}\par
Return only a single valid JSON object containing the above keys.
}
\end{aptdispbox}

\section{Coding Scheme}
\label{sec:coding_scheme}

Table~\ref{tab:coding_scheme} summarizes the coding scheme we used to analyze the 135 collected notation sketches. 
The scheme is organized by the three functional roles identified in our content analysis: \emph{geometric guides}, \emph{spatial guides}, and \emph{motion amplifiers}. 
We further subdivides each role into sub-categories (\eg, force, path, pose, style) with concrete code labels and brief operational definitions. 
During coding, annotators could assign any number of codes to a single drawing.

\newcommand{\vcell}[2]{\parbox[c]{#1}{\raggedright \vspace{1.1mm}
#2
\vspace{1.1mm}}}
\newcommand{\vcellc}[2]{\parbox[c]{#1}{\raggedright 
\vspace{1mm}
#2
\vspace{1mm}
}}

\newcommand{\WCat}{0.08\linewidth}
\newcommand{\WSub}{0.16\linewidth}
\newcommand{\WCode}{0.16\linewidth}
\newcommand{\WDef}{0.52\linewidth}

\begin{table*}[t]
\centering
\caption{The Coding scheme used for notation analysis for the 135 real-world collected drawings. During the coding, coders can assign any number of codes from this table to the same drawing.}
\label{tab:coding_scheme}
\footnotesize
\setlength{\tabcolsep}{4pt}

\begin{tabular}{l l l l l l}
\toprule
\vcellc{\WCat}{\textbf{Category}} &
\vcellc{\WSub}{\textbf{Sub-category}} &
\vcellc{\WCode}{\textbf{Code Label}} &
\vcellc{\WDef}{\textbf{Definition}} \\[5pt]
\midrule


\vcell{\WCat}{} &
\vcell{\WSub}{} &
\vcell{\WCode}{Overshoot} &
\vcell{\WDef}{Any marks indicating the object travels \textbf{past} its target destination before settling back.
} \\
\cline{3-4}

\vcell{\WCat}{} &
\vcell{\WSub}{} &
\vcell{\WCode}{Anticipation} &
\vcell{\WDef}{Any marks indicating a \textbf{preparatory action} in the \textbf{opposite} direction of the main action.} \\
\cline{3-4}

\vcell{\WCat}{} &
\vcell{\WSub}{Force (animation principles)} &
\vcell{\WCode}{Drag / Follow through} &
\vcell{\WDef}{Any marks on/near the \textbf{passive and loose parts} (e.g., hair, cloth, tails), showing them \textbf{lagging behind (drag)} or \textbf{swinging forward (follow through)} the main movement.} \\
\cline{3-4}

\vcell{\WCat}{} &
\vcell{\WSub}{} &
\vcell{\WCode}{Deformation (stretch/squash)} &
\vcell{\WDef}{Any marks that distort the shape to show flexibility: making the object \textbf{long and thin} for stretching, \textbf{short and wide} for squashing.} \\
\cline{2-4}

\vcell{\WCat}{} &
\vcell{\WSub}{} &
\vcell{\WCode}{Deformation (twist/tilt/bend)} &
\vcell{\WDef}{Any marks indicating the rotation of a body part \textbf{around its own axis} (e.g., hips, spine, neck) or bending of the subject, often for storing the energy.} \\
\cline{3-4}

\vcell{\WCat}{} &
\vcell{\WSub}{Force (actual forces)} &
\vcell{\WCode}{Gravity / Weight} &
\vcell{\WDef}{Any marks (often vertical lines, arrows pointing down) indicating the \textbf{downward pull or heaviness of an object}.} \\
\cline{3-4}

\vcell{\WCat}{} &
\vcell{\WSub}{} &
\vcell{\WCode}{Contact (Friction/Push off)} &
\vcell{\WDef}{Any marks at/near the point of interaction with a \textbf{surface} (e.g., feet on floor) indicating \textbf{frictions, launch force, or bounces}.} \\
\cline{3-4}

\vcell{\WCat}{} &
\vcell{\WSub}{} &
\vcell{\WCode}{Wind / Air flow} &
\vcell{\WDef}{Any marks surrounding the subject indicating \textbf{wind direction}.} \\
\cline{2-4}

\vcell{\WCat}{} &
\vcell{\WSub}{} &
\vcell{\WCode}{Strike Force (e.g., punch, hit, kick)} &
\vcell{\WDef}{Any marks directing a strike, punch, hit, usually \textbf{straight or sharp}.} \\
\cline{3-4}

\vcell{\WCat}{} &
\vcell{\WSub}{Force (action-specific)} &
\vcell{\WCode}{Rotational Force (e.g., swing / throw)} &
\vcell{\WDef}{Any marks indicating the \textbf{curved path of an object traveling through space} (e.g., a swing, a throw, a wide turn). Exclude if the object stays in roughly the same location; distinguish from twist/tilt.} \\
\cline{3-4}

\vcell{\WCat}{} &
\vcell{\WSub}{} &
\vcell{\WCode}{Action Vector (e.g., lift up)} &
\vcell{\WDef}{Linear arrows indicating the \textbf{direction of internal force, such as muscular effort} (e.g., the ``Lift up'', ``push off'').} \\
\cline{2-4}

\vcell{\WCat}{\textbf{Geometric guides}} &
\vcell{\WSub}{} &
\vcell{\WCode}{Line of action} &
\vcell{\WDef}{Broad, \textbf{sweeping primary curve(s)} on the subject that defines the \textbf{overall posture}, axis, and energy of the pose.} \\
\cline{3-4}

\vcell{\WCat}{} &
\vcell{\WSub}{Pose (living)} &
\vcell{\WCode}{Pose contour} &
\vcell{\WDef}{Multiple marks around the subject to \textbf{outline the contour of the subject}, often used to show how the subject's outer shape look like.} \\
\cline{3-4}

\vcell{\WCat}{} &
\vcell{\WSub}{} &
\vcell{\WCode}{Volumetric primitives} &
\vcell{\WDef}{Marks represent volumetric primitives (boxes, circles, triangles, etc.) to show \textbf{the scale of a specific part of the body}.} \\
\cline{2-4}

\vcell{\WCat}{} &
\vcell{\WSub}{} &
\vcell{\WCode}{Fabric folds} &
\vcell{\WDef}{Marks defining the stress points, tension, pinching, or bunching of \textbf{clothing/material}.} \\
\cline{3-4}

\vcell{\WCat}{} &
\vcell{\WSub}{Pose (non-living)} &
\vcell{\WCode}{Elemental form} &
\vcell{\WDef}{General \textbf{boundary shapes or flow lines} guiding groups of small amorphous elements (\textbf{smoke, fire, water, dust}).} \\
\cline{2-4}

\vcell{\WCat}{} &
\vcell{\WSub}{} &
\vcell{\WCode}{Trajectory (untimed)} &
\vcell{\WDef}{A line or multiple continuous marks tracing the \textbf{future path of movement}.} \\
\cline{3-4}

\vcell{\WCat}{} &
\vcell{\WSub}{Path} &
\vcell{\WCode}{Trajectory (timed)} &
\vcell{\WDef}{A path line that includes \textbf{small hash marks (ticks)} to indicate the exact \textbf{position of the object at each frame}.} \\
\cline{3-4}

\vcell{\WCat}{} &
\vcell{\WSub}{} &
\vcell{\WCode}{Directional guide} &
\vcell{\WDef}{Arrows showing which \textbf{direction} an object is facing or moving (NOT the full path trace).} \\
\cline{2-4}

\vcell{\WCat}{} &
\vcell{\WSub}{Style} &
\vcell{\WCode}{Motion strength} &
\vcell{\WDef}{Use of varying line weight (tapering or thickening strokes) or length to represent \textbf{intensity, speed}.} \\
\cline{3-4}

\vcell{\WCat}{} &
\vcell{\WSub}{} &
\vcell{\WCode}{Motion feeling} &
\vcell{\WDef}{Use of different style of the marks (e.g., different curveness) to indicate \textbf{how the animation feel}: e.g., it feels more bouncy/cartoony/fun vs.\ it feels more rigid/realistic.} \\
\hline

\vcell{\WCat}{} &
\vcell{\WSub}{Alignment arcs} &
\vcell{\WCode}{Alignment arcs} &
\vcell{\WDef}{Grid lines, horizontal/vertical guides, or ticks used to compare \textbf{relative sizes or align elements across frames}.} \\
\cline{2-4}

\vcell{\WCat}{
\vspace{-6mm}
\textbf{Spatial guides}} &
\vcell{\WSub}{Perspective guide} &
\vcell{\WCode}{Perspective guide} &
\vcell{\WDef}{Perspective grids, vanishing points, or receding lines used to indicate 3D depth and \textbf{size reduction}.} \\
\hline

\vcell{\WCat}{} &
\vcell{\WSub}{} &
\vcell{\WCode}{Motion lines} &
\vcell{\WDef}{Straight or arc strokes trailing \textbf{behind a mover} that indicate its path and suggest speed.} \\
\cline{3-4}

\vcell{\WCat}{} &
\vcell{\WSub}{} &
\vcell{\WCode}{Suppletion line} &
\vcell{\WDef}{Parts of the figure are replaced by streaks, creating a \textbf{blur that implies rapid motion}.} \\
\cline{3-4}

\vcell{\WCat}{
\textbf{Motion amplifiers}} &
\vcell{\WSub}{Smears} &
\vcell{\WCode}{Polymorphism} &
\vcell{\WDef}{Only a body part is \textbf{repeated} to show its movement through time.} \\
\cline{3-4}

\vcell{\WCat}{} &
\vcell{\WSub}{} &
\vcell{\WCode}{Backfixing line} &
\vcell{\WDef}{\textbf{Lines fixed to the background} (not the mover), creating a background blur that implies the camera is following the subject.} \\
\cline{2-4}

\vcell{\WCat}{} &
\vcell{\WSub}{Impact} &
\vcell{\WCode}{Circumfixing lines} &
\vcell{\WDef}{Short ticks ringing a figure or body part to show \textbf{vibration, wobble, or small repeated motion}.} \\
\cline{3-4}

\vcell{\WCat}{} &
\vcell{\WSub}{} &
\vcell{\WCode}{Impact star} &
\vcell{\WDef}{Starburst marks at the contact point to show the instant of \textbf{collision}.} \\
\bottomrule

\end{tabular}
\end{table*}

\newcommand{\plink}[1]{\texttt{\small #1}}
\newcommand{\NAWID}{0.05\textwidth}
\newcommand{\NAWLink}{0.28\textwidth}
\newcommand{\NAWCat}{0.10\textwidth}
\newcommand{\NAWSub}{0.22\textwidth}
\newcommand{\NAWLbl}{0.30\textwidth}

\section{Notation Analysis Details}
\label{sec:coding_result}

Table~\ref{tab:notation_analysis} provides an index of the 135 collected notation sketches, linking each example to its source and the codes assigned during our notation analysis. 
For each sketch, we provide its URL, along with the corresponding category, sub-category, and code label(s) from our coding scheme. 

\begin{table*}[t]
\centering
\caption{Notation Analysis Links with Labels.}
\label{tab:notation_analysis}
\footnotesize
\setlength{\tabcolsep}{3pt}
\renewcommand{\arraystretch}{1.15}

\begin{tabular}{r p{\NAWLink} p{\NAWCat} p{\NAWSub} p{\NAWLbl}}
\hline
\textbf{ID} & \textbf{Link} & \textbf{Category} & \textbf{Sub-category} & \textbf{Label} \\
\hline

1 & https://pin.it/5QbXhFnrv & Geometric & force (animation principle) & Deformation (stretch/squash) \\ \midrule
2 & https://pin.it/3NdkohLCa & Geometric & force (animation principle) & Deformation (stretch/squash) \\ \midrule
3 & https://pin.it/4GGkSeyqT & Geometric & force (animation principle) & Drag / Follow through \\ \midrule
4 & https://pin.it/7cEfXz5a1 & Geometric & force (actual force) & Gravity / Weight \newline Action vector \\ \midrule
5 & https://pin.it/7xT0kw3qB & Geometric & force (actual force) \newline force (action-specific ) & Gravity / Weight \newline Action vector \\ \midrule
6 & https://pin.it/2KbUF3gOB & Geometric & force (action-specific ) & Strike Force (e.g., punch, hit, kick) \newline Rotational Force (e.g., swing / throw) \\ \midrule
7 & https://pin.it/1fVJMv8Cy & Geometric & force (actual force) \newline force (action-specific ) & Gravity / Weight \newline Action vector \\ \midrule
8 & https://pin.it/1ZbFyGtU9 & Geometric & force (animation principle) \newline force (action-specific ) & Action vector \newline Overshoot \\ \midrule
9 & https://pin.it/7sSGiH0DR & Geometric & pose (living) & Line of action \\ \midrule
10 & https://pin.it/ApVBMmr2F & Geometric & pose (living) & Line of action \\ \midrule
11 & https://pin.it/35UVbJD2d & Geometric & pose (living) & Line of action \newline Pose contour \\ \midrule
12 & https://pin.it/78flBm3xD & Geometric & pose (living) & Line of action \\ \midrule
13 & https://pin.it/1hnaZoIkv & Geometric & pose (living) & Line of action \\ \midrule
14 & https://pin.it/1SToAr9ap & Geometric & pose (living) & Line of action \\ \midrule
15 & https://pin.it/vVLQFZMLe & Geometric & pose (living) & Line of action \\ \midrule
16 & https://pin.it/U4m0Yk2Pl & Geometric & pose (living) & Line of action \\ \midrule
17 & https://pin.it/5EalUN30B & Geometric & pose (living) & Line of action \\ \midrule
18 & https://pin.it/1bihRZfxY & Geometric & force (animation principle) & Overshoot \newline Anticipation \\ \midrule
19 & https://pin.it/132HdUAtQ & Geometric & force (action-specific ) \newline pose (living) & Action vector \newline Pose contour \\ \midrule
20 & https://pin.it/HVJXsewMU & Geometric & force (action-specific ) \newline pose (living) & Rotational Force (e.g., swing / throw) \newline Pose contour \newline Line of action \\ \midrule
21 & https://pin.it/5B5iMyWn9 & Geometric & force (action-specific ) \newline force (actual force) \newline force (animation principle) & Wind / Air flow \newline Action vector \newline Drag / Follow through \\ \midrule
22 & https://pin.it/3S9T5doki & Geometric & force (animation principle) & Deformation (twist/tilt/bend) \newline Deformation (stretch/squash) \\ \midrule
23 & https://pin.it/6ru01bCf6 & Geometric & style & Curved motion marks \\ \midrule
24 & https://pin.it/7wXYQDvrS & Geometric & pose (living) & Line of action \\ \midrule
25 & https://pin.it/O59kuS7xk & Geometric & pose (living) & Pose contour \\ \midrule
26 & https://pin.it/2vRm1Ewbh & Geometric & pose (non-living) & Elemental form \\ \midrule
27 & https://pin.it/7yogYNgSw & Geometric & pose (non-living) & Fabric folds \\ \midrule
28 & https://pin.it/cbz5ibJkR & Geometric & pose (non-living) & Elemental form \\ \midrule
29 & https://pin.it/79iafwYmS & Geometric & pose (non-living) & Elemental form \\ \midrule
30 & https://pin.it/7fo2nztk0 & Geometric & pose (living) \newline force (animation principle) & Deformation (twist/tilt/bend) \newline Line of action \\ \midrule
31 & https://pin.it/7ziPTLQEy & Geometric & force (actual force) \newline pose (living) & Gravity / Weight \newline Pose contour \\ \midrule
32 & https://pin.it/4zfu3IQ0v & Geometric & pose (living) & Line of action \newline Volumetric primitives \\ \midrule
33 & https://pin.it/4cVkHojkf & Geometric & pose (living) \newline force (action-specific ) & Line of action \newline Action vector \\ \midrule
34 & https://pin.it/1GWIGRShf & Geometric & pose (living) & Line of action \\ \midrule
35 & https://pin.it/4HKx9LCB2 & Geometric & pose (living) \newline force (animation principle) & Pose contour \newline Deformation (stretch/squash) \\ \midrule
 \aptLtoX[graphic=no,type=html]{}{& & & &  Continue on the next page....\\
 \midrule
\end{tabular}
\end{table*}
\begin{table*}[t]
\centering
\caption{Notation Analysis Links with Labels (continued).}
\footnotesize
\setlength{\tabcolsep}{3pt}
\renewcommand{\arraystretch}{1.15}
\begin{tabular}{r p{\NAWLink} p{\NAWCat} p{\NAWSub} p{\NAWLbl}}
\hline
\textbf{ID} & \textbf{Link} & \textbf{Category} & \textbf{Sub-category} & \textbf{Label} \\
\hline
}
36 & https://pin.it/1tNanb2J5 & Geometric & path & Trajectory with spacing ticks \\ \midrule
37 & https://pin.it/2AMkUcm77 & Geometric & path & Directional guide \\ \midrule
38 & https://pin.it/3QH8DCVJC & Geometric & path \newline pose (non-living) & Directional guide \newline Elemental form \\ \midrule
39 & https://pin.it/13h9TxWGl & Geometric & pose (living) & Line of action \\ \midrule
40 & https://pin.it/bm33gOPn7 & Geometric & pose (living) & Line of action \newline Pose contour \\ \midrule
41 & https://pin.it/3KBbe9lkf & Geometric & pose (living) & Volumetric primitives \\ \midrule
42 & https://pin.it/3EaqDl1cx & Geometric & pose (living) & Line of action \\ \midrule
43 & https://pin.it/1XbSboKH9 & Geometric & pose (living) \newline force (action-specific ) & Line of action \newline Rotational Force (e.g., swing / throw) \\ \midrule
44 & https://pin.it/1RV910dNn & Geometric & force (actual force) & Deformation (twist/tilt/bend) \\ \midrule
45 & https://pin.it/OurMFYWr6 & Geometric & force (action-specific ) & Action vector \\ \midrule
46 & https://pin.it/3RkEmRbUf & Geometric & force (action-specific ) \newline force (animation principle) & Strike Force (e.g., punch, hit, kick) \newline Anticipation \\ \midrule
47 & https://pin.it/4Q6ioCBOA & Geometric & pose (living) & Line of action \newline Volumetric primitives \\ \midrule
48 & https://pin.it/1bEv0WrLq & Geometric & pose (non-living) & Fabric folds \\ \midrule
49 & https://pin.it/K7Zi2K12t & Geometric & force (animation principle) \newline path & Deformation (stretch/squash) \newline Trajectory \\ \midrule
50 & https://pin.it/4LAJF2OTc & Geometric & path & Trajectory \\ \midrule
51 & https://pin.it/2PXqLBUEK & Geometric & force (action-specific ) \newline force (actual force) & Deformation (twist/tilt/bend) \newline Rotational Force (e.g., swing / throw) \\ \midrule
52 & https://pin.it/4spjPxyPR & Geometric & pose (living) & Line of action \\ \midrule
53 & https://pin.it/1pm1yARNr & Geometric & force (action-specific ) & Rotational Force (e.g., swing / throw) \newline Strike Force (e.g., punch, hit, kick) \\ \midrule
54 & https://pin.it/3V1Vx7OkS & Geometric & pose (living) & Line of action \\ \midrule
55 & https://pin.it/1YCJBWguK & Geometric & pose (living) \newline force (action-specific ) & Line of action \newline Action vector \newline Strike Force (e.g., punch, hit, kick) \\ \midrule
56 & https://pin.it/2EbnyUmPD & Geometric & pose (living) & Line of action \\ \midrule
57 & https://pin.it/puVzdr7gd & Geometric & pose (living) & Line of action \\ \midrule
58 & https://pin.it/1HUvtwUQ6 & Geometric & path & Directional guide \\ \midrule
59 & https://pin.it/3nDQGrt8A & Geometric & path & Directional guide \\ \midrule
60 & https://pin.it/7ITbqxkG8 & Geometric & pose (living) \newline force (animation principle) & Deformation (stretch/squash) \newline Pose contour \\ \midrule
61 & https://pin.it/3CKDxV5Eg & Geometric & path & Trajectory \newline Directional guide \\ \midrule
62 & https://pin.it/40YbnR2GI & Geometric & pose (living) & Pose contour \\ \midrule
63 & https://pin.it/4Yj1PVyv7 & Geometric & pose (living) & Line of action \\ \midrule
64 & https://pin.it/22GqWHFMc & Geometric & pose (living) & Pose contour \\ \midrule
65 & https://pin.it/30jX4VYJ9 & Geometric & force (animation principle) \newline force (actual force) & Drag / Follow through \newline Gravity / Weight \newline Wind / Air flow \\ \midrule
66 & https://pin.it/3hdQn7oPT & Geometric & force (action-specific ) \newline pose (living) \newline path & Rotational Force (e.g., swing / throw) \newline Contact (Friction/Push off) \newline Action vector \newline Trajectory with spacing ticks \\ \midrule
67 & https://pin.it/1uhVsRKOc & Geometric & force (actual force) \newline force (action-specific ) \newline pose (living) & Gravity / Weight \newline Action vector \newline Volumetric primitives \\ \midrule
68 & https://pin.it/2rmhc9hOy & Geometric & force (actual force) \newline pose (living) & Strike Force (e.g., punch, hit, kick) \newline Pose contour \\ \midrule
69 & https://pin.it/24LKnlyGt & Geometric & style & Curved motion marks \\ \midrule
70 & https://pin.it/6SrjzpA3E & Geometric & pose (living) \newline force (action-specific ) & Line of action \newline Rotational Force (e.g., swing / throw) \\ \midrule
\aptLtoX[graphic=no,type=html]{}{ & & & &  Continue on the next page....\\
 \midrule

\end{tabular}
\end{table*}
\begin{table*}[t]
\centering
\caption{Notation Analysis Links with Labels (continued).}
\footnotesize
\setlength{\tabcolsep}{3pt}
\renewcommand{\arraystretch}{1.15}
\begin{tabular}{r p{\NAWLink} p{\NAWCat} p{\NAWSub} p{\NAWLbl}}
\hline
\textbf{ID} & \textbf{Link} & \textbf{Category} & \textbf{Sub-category} & \textbf{Label} \\
\hline
}
71 & https://pin.it/5f1ZEUeGW & Geometric & pose (living) & Pose contour \\ \midrule
72 & https://pin.it/ZYY33KA9O & Geometric & pose (non-living) & Fabric folds \\ \midrule
73 & https://pin.it/FGKEx7ObN & Geometric & pose (non-living) & Fabric folds \\ \midrule
74 & https://pin.it/3Sa2g1a3e & Geometric & force (animation principle) \newline path \newline force (action-specific ) & Deformation (stretch/squash) \newline Action vector \newline Directional guide \\ \midrule
75 & https://pin.it/5TBehyCwR & Geometric & force (actual force) \newline path & Contact (Friction/Push off) \newline Trajectory \\ \midrule
76 & https://pin.it/7aIC3dggL & Geometric & pose (non-living) & Fabric folds \\ \midrule
77 & https://pin.it/31TTy0le3 & Geometric & pose (non-living) & Elemental form \\ \midrule
78 & https://pin.it/6q0ftCUNW & Geometric & force (animation principle) \newline force (actual force) & Deformation (stretch/squash) \newline Deformation (twist/tilt/bend) \newline Wind / Air flow \\ \midrule
79 & https://pin.it/1FZBIo8Os & Geometric & force (action-specific ) \newline pose (living) & Action vector \newline Pose contour \\ \midrule
80 & https://pin.it/27k1wWgQa & Geometric & force (action-specific ) & Rotational Force (e.g., swing / throw) \newline Strike Force (e.g., punch, hit, kick) \\ \midrule
81 & https://pin.it/5vIFPlyJM & Geometric & force (actual force) \newline force (action-specific ) & Wind / Air flow \newline Action vector \\ \midrule
82 & https://pin.it/29wFA4jDK & Geometric & force (animation principle) & Deformation (stretch/squash) \\ \midrule
83 & https://pin.it/5LKNGN5bc & Geometric & pose (living) & Line of action \\ \midrule
84 & https://pin.it/5aotIp3ZC & Geometric & pose (living) & Pose contour \\ \midrule
85 & https://pin.it/5oYpm9kzZ & Geometric & force (actual force) \newline pose (living) & Deformation (twist/tilt/bend) \newline Pose contour \\ \midrule
86 & https://pin.it/3m0tYgjwz & Geometric & force (animation principle) \newline force (action-specific ) & Drag / Follow through \newline Action vector \\ \midrule
87 & https://pin.it/452hndcKi & Geometric & path & Directional guide \\ \midrule
88 & https://pin.it/2rkAKgXmT & Geometric & force (animation principle) \newline force (action-specific ) & Drag / Follow through \newline Action vector \\ \midrule
89 & https://pin.it/52ZhA5el9 & Geometric & pose (living) & Volumetric primitives \\ \midrule
90 & https://pin.it/7uAmm5uu5 & Geometric & force (animation principle) \newline force (action-specific ) \newline force (actual force) & Deformation (stretch/squash) \newline Gravity / Weight \newline Action vector \\ \midrule
91 & https://pin.it/barxlr0Ig & Geometric & path \newline force (animation principle) & Deformation (stretch/squash) \newline Trajectory \\ \midrule
92 & https://pin.it/4rhjWTOTD & Geometric & path & Directional guide \\ \midrule
93 & https://pin.it/6vEKiGYQd & Geometric & pose (living) \newline force (action-specific ) & Action vector \newline Line of action \\ \midrule
94 & https://pin.it/49AoEF9E3 & Geometric & force (action-specific ) \newline pose (living) & Action vector \newline Volumetric primitives \newline Line of action \\ \midrule
95 & https://pin.it/65A5OmwQt & Geometric & force (action-specific ) \newline pose (living) & Line of action \newline Action vector \\ \midrule
96 & https://pin.it/EfIyVorZ8 & Geometric & force (action-specific ) & Action vector \newline Rotational Force (e.g., swing / throw) \\ \midrule
97 & https://pin.it/3qHZQreXG & Geometric & path & Directional guide \\ \midrule
98 & https://pin.it/2zfsVedJP & Geometric & path & Directional guide \\ \midrule
99 & https://pin.it/1M7OAlt1F & Geometric & path \newline force (actual force) & Trajectory \newline Contact (Friction/Push off) \\ \midrule
100 & https://pin.it/5T9MbSeLA & Geometric & path & Trajectory with spacing ticks \\ \midrule
101 & https://pin.it/1gyN7HyFm & Geometric & path & Trajectory \\ \midrule
102 & https://pin.it/7hLEmwODJ & Geometric & path & Trajectory \\ \midrule
103 & https://pin.it/6I0kmI6o4 & Spatial & perspective guide & Size reduction guide \\ \midrule
104 & https://pin.it/1uGIl9IP9 & Spatial & perspective guide & Size reduction guide \\ \midrule
105 & https://pin.it/4JHb9YZ6L & Spatial & perspective guide & Size reduction guide \\ \midrule
\aptLtoX[graphic=no,type=html]{}{
 & & & &  Continue on the next page....\\
 \midrule
\end{tabular}
\end{table*}
\begin{table*}[t]
\centering
\caption{Notation Analysis Links with Labels (continued).}
\footnotesize
\setlength{\tabcolsep}{3pt}
\renewcommand{\arraystretch}{1.15}
\begin{tabular}{r p{\NAWLink} p{\NAWCat} p{\NAWSub} p{\NAWLbl}}
\hline
\textbf{ID} & \textbf{Link} & \textbf{Category} & \textbf{Sub-category} & \textbf{Label} \\
\hline}
106 & https://pin.it/6BcGF6zWu & Spatial & perspective guide & Size reduction guide \\ \midrule
107 & https://pin.it/6zyEQa6tI & Spatial & perspective guide & Size reduction guide \\ \midrule
108 & https://pin.it/4Q2bWwdH1 & Spatial & perspective guide & Size reduction guide \\ \midrule
109 & https://pin.it/1yRNEyVi5 & Spatial & Alignment arcs & Porportion alignment \\ \midrule
110 & https://pin.it/6LHTJu0j7 & Spatial & Alignment arcs & Alignment arcs \\ \midrule
111 & https://pin.it/wfKjtApIa & Spatial & Alignment arcs & Alignment arcs \\ \midrule
112 & https://pin.it/bRpfObjCR & Spatial & Alignment arcs & Alignment arcs \\ \midrule
113 & https://pin.it/345tQpw2U & Spatial & Alignment arcs & Alignment arcs \\ \midrule
114 & https://pin.it/2JvHErrWL & Spatial & Alignment arcs & Alignment arcs \\ \midrule
115 & https://pin.it/6wD13GnYT & Spatial & Alignment arcs & Alignment arcs \\ \midrule
116 & https://pin.it/4mErVjNSB & Spatial & Alignment arcs & Alignment arcs \\ \midrule
117 & https://pin.it/7ipV05HHg & Spatial & Alignment arcs & Alignment arcs \\ \midrule
118 & https://pin.it/2AEpzmYXf & Spatial & Alignment arcs & Alignment arcs \\ \midrule
119 & https://pin.it/3obYNXpaD & Spatial & Alignment arcs & Alignment arcs \\ \midrule
120 & https://pin.it/4dZzRFntn & Spatial & Alignment arcs & Alignment arcs \\ \midrule
121 & https://pin.it/3z3IEwa9S & Spatial & Alignment arcs & Alignment arcs \\ \midrule
122 & https://pin.it/4q7c0dJVp & Spatial & Alignment arcs & Alignment arcs \\ \midrule
123 & https://pin.it/493T8KgaW & Spatial & Alignment arcs & Alignment arcs \\ \midrule
124 & https://pin.it/bhsKH21R0 & Spatial & Alignment arcs & Alignment arcs \\ \midrule
125 & https://pin.it/qpdmQr5Dh & Amplifier & impact & Impact star \\ \midrule
126 & https://pin.it/19S05gaFo & Amplifier & smears & Suppletion line \\ \midrule
127 & https://pin.it/4Fp0jGKv2 & Amplifier & impact & Impact star \\ \midrule
128 & https://pin.it/LYBv2V0h4 & Amplifier & smears & Polymorphism \\ \midrule
129 & https://pin.it/3XcFsLmGp & Amplifier & smears & Polymorphism \\ \midrule
130 & https://pin.it/4D6CnOVF8 & Amplifier & impact & Impact star \\ \midrule
131 & https://pin.it/6G7SVJK7U & Amplifier & impact & Circumfixing lines \\ \midrule
132 & https://pin.it/6iLOulISY & Amplifier & smears & Backfixing line \\ \midrule
133 & https://pin.it/1bOmi5kKs & Amplifier & smears & Motion lines \\ \midrule
134 & https://pin.it/TO5sCVqhv & Amplifier & smears & Backfixing line \\ \midrule
135 & https://pin.it/7lrjeTPoh & Amplifier & smears & Motion lines \\ \bottomrule

\end{tabular}
\end{table*}
\section{VLM Responses with and without Structured Representation}
\label{sec:vlm_responses}

\begin{figure}[htp]
    \includegraphics[width=\linewidth]{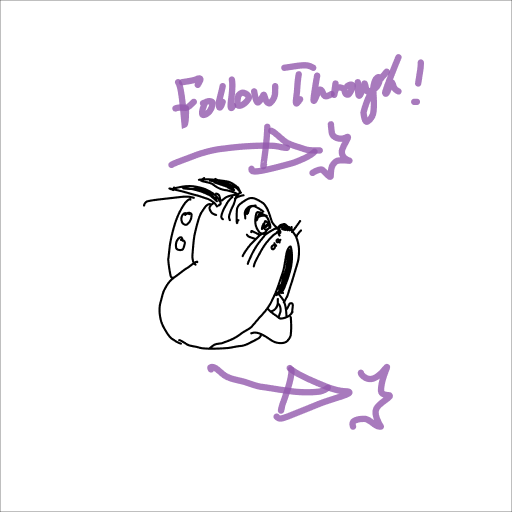}
    \caption{The test case to interpret with and w/o the structured representation.}
    \label{fig:vlm_response_case}
\end{figure}

In this section, we present a qualitative comparison between VLM responses generated with and without our structured representation, interpreting the case of Figure~\ref{fig:vlm_response_case}. 
While the unstructured baseline correctly identifies the general context of a head movement, it fails to capture the key essence of how the animation principle, follow through, applies to the soft ear and cheeks of the character instead of the whole head of the character. 
Specifically, it aggregates the distinct movements of the ears and jowls into a single ``gentle S-curve'' trajectory, treating the head as a rigid body. 
It overlooks the critical temporal and spatial separation between the loose upper and lower masses, losing the nuance of the overlapping action that defines the follow-through effect.

\vspace{5mm}

\begin{aptdispbox}
\acmcallout{VLM's response w/o structured representation}{
The sketch is giving a note for ``follow-through'' on the bulldog’s head move.
The bulldog snaps its head toward the right.  
After the main snap, the head should keep traveling along the same right-ward arc, dipping slightly downward, then settle.  
The path of action is indicated by two curved arrows that form a gentle S-curve --- first a small upward-right swing, then a broader downward-right sweep.  
Little star-shaped burst marks at the arrow tips suggest a soft ease-out / settling at the end of the arc.
}\end{aptdispbox}

\vspace{8mm}

\begin{aptdispbox}
\acmcallout{VLM's response with structured representation}{

\begin{itemize}[leftmargin=*]
    \item \textbf{UNIT 1}
    \begin{itemize}[leftmargin=*]
        \item \textbf{Primary Components}
        \begin{itemize}
            \item \textbf{Source:} Loose upper portions of the dog's head (ear tips / top-tuft).
            \item \textbf{Path:} Broad, slightly upward-curving arrow that sweeps left-to-right over the head.
            \item \textbf{Target:} Positions a little farther to camera-right, continuing after the main head stop (follow-through).
        \end{itemize}
        
        \item \textbf{Secondary Modifiers}
        \begin{itemize}[leftmargin=*]
            \item \textbf{Color:} Purple \textit{(Meaning: motion cue, Scope: unit)}
            \item \textbf{Thickness:} Medium \textit{(Meaning: principal path, Scope: path)}
            \item \textbf{Text:} ``Follow Through!'' \textit{(Meaning: clarify that the motion is a follow-through action, Scope: unit)}
            \item \textbf{Symbol:} Small 4-point star at arrow tip \textit{(Meaning: accent / settle point, Scope: target)}
        \end{itemize}
        
        \item \textbf{Temporal Order:} 1
        \item \textbf{Natural Language Summary:} After the head stops, the floppy ear and top mass keep sliding a little farther to the right, illustrating the follow-through principle.
    \end{itemize}

    \vspace{0.5cm} 

    \item \textbf{UNIT 2}
    \begin{itemize}[leftmargin=*]
        \item \textbf{Primary Components}
        \begin{itemize}
            \item \textbf{Source:} Loose lower portions of the head (cheeks/jowls \& collar tag).
            \item \textbf{Path:} Gentle left-to-right, slightly downward arrow arcing beneath the chin.
            \item \textbf{Target:} Overshoots to a lowered, right-shifted settle position.
        \end{itemize}
        
        \item \textbf{Secondary Modifiers}
        \begin{itemize}[leftmargin=*]
            \item \textbf{Color:} Purple \textit{(Meaning: motion cue, Scope: unit)}
            \item \textbf{Thickness:} Medium \textit{(Meaning: principal path, Scope: path)}
            \item \textbf{Symbol:} Small 4-point star at arrow tip \textit{(Meaning: accent / settle point, Scope: target)}
        \end{itemize}
        
        \item \textbf{Temporal Order:} 2
        \item \textbf{Natural Language Summary:} Simultaneously or a beat later, the heavy jowls and collar tag lag and then swing past the stopping point of the skull, settling lower and farther to the right.
    \end{itemize}
\end{itemize}

}
\end{aptdispbox}

\end{document}